\documentclass[12pt]{article}

\usepackage{latexsym}
\usepackage{amsthm}
\usepackage{amsmath}
\usepackage[dvips]{graphicx}
\usepackage{amssymb}
\usepackage{amsbsy}

\newcommand{\be}{\begin{eqnarray}}
\newcommand{\ee}{\end{eqnarray}}

\newcommand{\MM}{\mathcal{M}}
\newcommand{\ZZ}{\mathcal{Z}}
\newcommand{\WW}{\mathcal{W}}
\newcommand{\LL}{\mathcal{L}}

\begin{document}

\baselineskip=18pt

\setcounter{footnote}{0}
\setcounter{figure}{0}
\setcounter{table}{0}

\begin{titlepage}
\vspace{1cm}

\begin{center}

{\Large \bf Unraveling ${\cal L}_{n,k}$: Grassmannian Kinematics}

\vspace{0.8cm}

{\bf Jared Kaplan\footnote{jaredk@slac.stanford.edu}}

\vspace{.5cm}

{\it Theory Group, SLAC National Accelerator Laboratory, Menlo Park, CA 94025, USA}

\end{center}

\vspace{1cm}

\begin{abstract}

It was recently proposed that the leading singularities of the S-Matrix of ${\cal N} = 4$ super Yang-Mills theory arise as the residues of a contour integral over a Grassmannian manifold, with space-time locality encoded through residue theorems generalizing Cauchy's theorem to more than one variable.  We provide a method to identify the residue corresponding to any leading singularity, and we carry this out explicitly for all leading singularities at tree level and one-loop.  We also give several examples at higher loops, including all generic two-loop leading singularities and an interesting four-loop object.  As an example we consider a $12$-pt N$^4$MHV leading singularity at two loops that has a kinematic structure involving double square roots.  Our analysis results in a simple picture for how the topological structure of loop graphs is reflected in various substructures within the Grassmannian.

\end{abstract}

\bigskip
\bigskip

\end{titlepage}

\section{Introduction and Review}

A proposal was recently made that all of the leading singularities of ${\cal N} = 4$ super Yang-Mills theory in the large $N$ limit arise as the residues of a contour integral over a Grassmannian manifold \cite{ArkaniHamed:2009dn}.  It has been conjectured that these leading singularities may be sufficient to determine the perturbative S-Matrix of the theory \cite{Cachazo:2008vp},\cite{simplest}, and this has been confirmed for all one-loop amplitudes \cite{Bern:1993mq}-\cite{Bern:2004ky} and for a few simple examples at higher loops \cite{Cachazo:2008hp}-\cite{zzbb}.  Thus it is hoped that this strikingly new portrayal of the S-Matrix may be part of a new description of scattering, where the extreme simplicity of the S-Matrix itself takes center stage and space-time locality is encoded in a complicated way.  

The Grassmannian contour integral was discovered through investigations \cite{Hodges:2005bf}-\cite{Korchemsky:2009jv}
 of scattering amplitudes and the BCFW Recursion Relations \cite{bcf}-\cite{Drummond:2008cr}
in twistor space \cite{Penrose:1967wn}-\cite{Penrose:1999cw},
 inspired in part by the twistor string \cite{Witten:2003nn}, but it remains a mysterious new object without any clear first-principled derivation.  The case for its validity was based on two sources of evidence, in addition to the fact that it possess all of the required symmetries \cite{ArkaniHamed:2009dn}, \cite{ArkaniHamed:2009vw}, \cite{Mason:2009qx}, including dual conformal invariance \cite{Alday:2007hr}-\cite{McGreevy:2008zy}.   The first piece of evidence was the explicit computation of various residues and their subsequent identification among known leading singularities \cite{ArkaniHamed:2009dn}, \cite{Mason:2009qx}.  The second and perhaps more interesting piece of evidence was based on an analysis of the residue theorems that follow from generalizations of Cauchy's theorem to more than one variable.  It was shown in many examples \cite{ArkaniHamed:2009dn} that these residue theorems are directly related to space-time locality, as they enforce the cancellation of unphysical poles in and the symmetries of tree amplitudes and the Infrared consistency of one-loop amplitudes.  Some of these residue theorems imply non-trivial relations that do not follow from the one-loop IR equations \cite{IReq} and that were conjectured to follow instead from IR consistency at higher loops.

\begin{figure}[htp]
\begin{center}
\includegraphics[width=15cm]{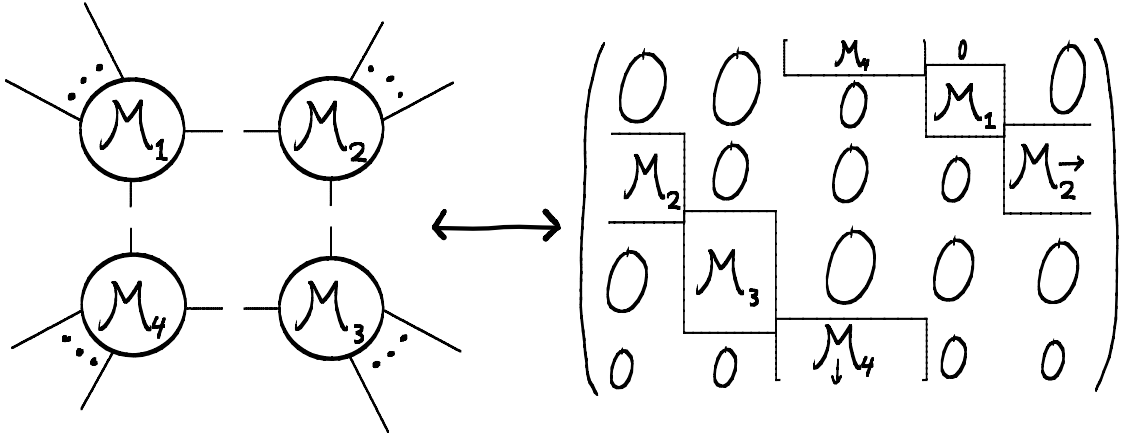} 
\caption{We illustrate the way that one-loop leading singularities correspond to certain subspaces of the Grassmannian.  The object on the left is a one-loop leading singularity, or in other words it is the product of four tree amplitudes evaluated on the kinematics determined by the quadruple cut of the loop integral.  The rectangles in the picture on the right are the non-zero entries of the $k \times n$ matrix characterizing the Grassmannian; each rectangular block shares one row with the block adjacent to it.  It should be noted that only $GL(k)$ invariant statements about this matrix are physically meaningful.} \label{BoxasMatrix}
\end{center}
\end{figure}

The purpose of the present paper is to provide a simple picture for how leading singularities emerge as the residues of the Grassmannian contour integral, which we will refer to as $\LL_{n,k}$.  Our methods allow us to identify a residue of $\LL_{n,k}$ corresponding to any given leading singularity.  We will carry out this procedure explicitly at tree level and at one-loop, and give a few illustrative examples at higher loops.  Our analysis will be `kinematical' as opposed to `dynamical' in a sense that will be made clear below, so we will not actually prove that every leading singularity is in fact a residue, but we believe our analysis is nevertheless very powerful.  We find it especially striking that the topological structure of the loop graph corresponding to a given leading singularity is reflected in the structure of the Grassmannian; this can be seen already in figure \ref{BoxasMatrix}.

Leading singularities and Grassmannian contour integrals are not widely known, so we will briefly review both.  The computation of scattering amplitudes in terms of their leading singularities is a descendent of generalized unitarity techniques \cite{Dixon:1996wi}.  As will be familiar from Feynman diagram computations, loop amplitudes involve various logarithms, dilogarithms, and so on that are themselves functions of the kinematical invariants of the scattering process.  These functions have branch cuts, and one can compute the discontinuities across these cuts.  Those discontinuities may themselves have branch cuts, and we can compute these discontinuities, and so on, until we are left with some pure rational functions (we get many different rational functions depending on which branch cuts we use, and which loop order we are at).  These rational functions are the leading singularities of a scattering amplitude, and it has been conjectured \cite{simplest} that the leading singularities are sufficient information to reconstruct the S-Matrices of ${\cal N} = 4$ SYM and ${\cal N} = 8$ Supergravity.

At this point the leading singularity may seem like a rather technical construction, but in fact it is a simple and physical object.  The reason is that the branch cut of an integral (such as a loop integral) is approached when a parameter in the integrand forces the contour of integration to encircle a pole.  In a local quantum field theory, poles in the loop integrand can only come from propagators, so by isolating the discontinuity across a branch cut we are forcing the virtual particles in the loops to go on-shell.  Leading singularities arise when all of the loop integrations are fixed (or `cut') by the requirement that various intermediate particles are on-shell.  Thus leading singularities are simply products of tree-level scattering amplitudes evaluated with very special kinematical configurations.  If the full S-Matrix is determined by leading singularities, then it is determined by the classical scattering amplitudes of the theory in the simplest possible way.  

Now let us describe our Grassmannian contour integral.  A Grassmannian manifold $G(k,n)$ is the space of $k$ dimensional planes in an $n$ dimensional space.  A convenient way to parameterize the points of $G(k,n)$ is with a $k \times n$ matrix $C_{\alpha a}$, where $\alpha = 1,...,k$ and $a = 1,...,n$; the rows of this matrix span a $k$ plane.  Note that different $C$ matrices related by a $GL(k)$ transformation $C_{\alpha a} \to L_\alpha^{\ \beta} C_{\beta a}$  correspond to the same $k$-plane, so $GL(k)$ is a ``gauge symmetry'' of our description of the Grassmannian.

In what follows the parameter $n$ will always correspond to the number of particles in a scattering amplitude or leading singularity, and $k$ will represent the total number of negative helicity gluons in an all-gluon amplitude (or more generally the R-charge sector), so MHV amplitudes \cite{Parke:1986gb} correspond to $k=2$.  The contour integral we will consider is an integral over the $C$ matrices with a very special integrand:
\be
\LL_{n,k}({\cal W}_a) = \int \frac{d^{k \times n} C_{\alpha
a}}{(12\cdots k) \, (23\cdots (k+1) \,) \, \cdots (n 1 \cdots (k-1)
\,)} \prod_{\alpha = 1}^k \delta^{4|4}(C_{\alpha a} {\cal W}_a) 
\ee
The factors in the denominator are the determinants of the sequential $k \times k$ minors of $C$, explicitly they are 
\be
(m_1 ... m_k) = \epsilon^{\alpha_1 ... \alpha_k} C_{m_1 \alpha_1} ... C_{m_k \alpha_k}
\ee
The other piece of the integrand is a product of $k$ superconformal delta functions, and this is where the dependence on the kinematic variables of the external particles enters.  We represent the kinematics with twistor variables $\WW$ where
\be
\WW = (\tilde \lambda, \tilde \mu, \tilde \eta) 
\ee
and $\tilde \mu$ is the Fourier conjugate to the spinor variable $\tilde \lambda$, with $p_\mu = \lambda \sigma_\mu \tilde \lambda$.  Note that these super twistor variables $\WW$ are in the fundamental representation of the superconformal group $PSU(2,2|4)$.  The anti-commuting $\tilde \eta$ variable is an on-shell superspace coordinate \cite{Nair:1988bq}.  The use of twistor variables for scattering amplitudes has been extensively and pedagogically discussed in \cite{ArkaniHamed:2009si}, and on-shell superspace in \cite{simplest}; we will not review them further here.

To begin to better understand $\LL_{n,k}$ let us count the number of integration variables in momentum space.  To go to momentum space we just Fourier transform with respect to the $\mu_a$ variables, giving
\be
\LL_{n,k}(\lambda, \tilde \lambda, \eta) = \int \frac{d^{k \times n} C_{\alpha
a} \ d^{2k} \rho_\alpha \ \ \prod_{\alpha=1}^k (C_{\alpha a} \tilde \eta_a)^4 }{(12\cdots k) \, (23\cdots (k+1) \,) \, \cdots (n 1 \cdots (k-1) \,)} \delta^{2k}(C_{\alpha a} \tilde \lambda_a) \delta^{2n}(\lambda_a - C_{\alpha a}  \rho_\alpha)
\ee
where the $\rho_\alpha$ are extra spinor variables to be integrated over.
We see that after eliminating these extra spinors there are $2n$ delta functions, but $4$ of these encode momentum conservation.  This means that $2n-4$ of the coordinates in the $C_{\alpha a}$ matrix will be fixed by these delta functions.  Also, some $k^2$ of the coordinates can be eliminated by fixing the $GL(k)$ gauge redundancy of the Grassmannian.  All of the remaining $(n-k-2)(k-2)$ coordinates are free, so $\LL_{n,k}$ should be regarded as a contour integral in this many variables.  The choice of contour or residue can be viewed as providing equations that fix the integration variables, but we can perform the contour integral and solve the delta function constraints in whatever order we prefer.  In \cite{ArkaniHamed:2009dn} we solved the delta function constraints first, and only then performed the contour integration, but we will find the opposite order to be more enlightening in what follows\footnote{One might worry that there exist contours of integration that are incompatible with the delta function constraints.  We will never be led to such `bad' contours, although they are a reasonable motivation for solving the delta function constraints before performing the contour integration.}.

Once the contour integration is performed so that we are left with one particular residue, the full Grassmannian will be reduced to some $2n-4$ dimensional algebraic subspace parameterized by a highly constrained $C_{\alpha a}$ matrix.  As a very concrete example that we will derive below, the matrix
\be
C = \left(\begin{array}{cccccccc} 
c_{21} & 1 & 0 & 0 & 0 & 0 & c_{27} & c_{28} \\ 
c_{41} & c_{42} & c_{43} & 1 & 0 & 0 & 0 & 0 \\
0 & 0 & c_{63} & c_{64} & c_{65} & 1 & 0 & 0 \\
0 & 0 & 0 & 0 & c_{85} & c_{86} & c_{87} & 1 
\end{array} \right) 
\ee
corresponds to a one-loop leading singularity with a $4$-pt MHV amplitude at each of the four corners of the `box' pictured in figure \ref{BoxasMatrix}.  This is a rather remarkable result, because it means that all leading singularities essentially only depend on kinematic invariants through $2n-4$ special parameters, whereas we might expect them to depend on the $n(n-1)$ invariants $\langle i j \rangle$ and $[i j]$.  This is especially surprising when we remember that this is an ${\cal N} = 4$ supersymmetric result, so it holds for all of the various helicity combinations.

The methods we will develop in the following sections will allow us to pick out the special subspaces within the Grassmannian that give rise to any given leading singularity.  We will show that there is a very simple way to glue together many smaller copies of $\LL_{n,k}$ so that they sit as subspaces of a larger Grassmannian, where the smaller copies are to be interpreted as tree amplitudes (or general leading singularities) at the vertices of a loop diagram that has been `cut' to make a larger leading singularity.  

Once we have identified an appropriate subspace within the larger Grassmannian, there still remains the question of whether this subspace can actually be obtained as a residue.  We show that this is extremely plausible in the appendix.  However, our analysis is `kinematical' as opposed to `dynamical' because we are not able to actually compute these residues in general.  A full proof that all leading singularities are residues of $\LL_{n,k}$ would require this computation, and this is beyond the scope of the present work.

In the next section we show how leading singularities can be written in twistor space, and in particular how they can be computed by `gluing' together other leading singularities.  Then in section three we begin by motivating our analysis, and then we proceed to identify all tree and one-loop leading singularities.  At the end of section three we give some very non-trivial higher loop examples, including all generic two-loop leading singularities and a four loop object with an interesting topological structure (as a loop graph).  Also, to show the power of our method we provide an explicit $12$-pt N$^4$MHV two-loop example whose kinematic structure involves square roots of square roots.   With section four we conclude and discuss future directions.  In an appendix we give some details of the computation of the residues themselves, including an  argument for the existence of the tree and one-loop residues, and we give an explicit solution for the NMHV ($k=3$) sector.

\section{Leading Singularities in Twistor Space}

Twistor variables are an elegant representation of massless on-shell states, so phase space integrals such as
\be
\int d^4 \ell \delta(\ell^2) M_1(\ell) M_2(-\ell)
\ee
can be written very simply in twistor space as
\be
\int D^3 W_P M_1(W_P) M_2(W_P)
\ee
This is an instance of the well-known Penrose transform \cite{Penrose:1967wn}.  It is essentially guaranteed by Lorentz invariance and the kinematics of twistor space -- in other words, since twistors fully parameterize light-like states, what else could an integral over twistor space be but a dLIPS integral -- but let us derive the result explicitly.

We begin by recalling that the momentum vector $\ell_\mu$ can be written in spinor language as the $2 \times 2$ matrix 
\be
\ell_\mu \sigma^\mu_{\alpha \dot \alpha} = \left( \begin{array}{cc} \ell^+ & \ell^\perp \\ \tilde \ell^\perp & \ell^- \end{array} \right)
\ee
We will use $(2,2)$ signature to facilitate calculation, but all of the results we will obtain can be analytically continued back to the usual $(3,1)$ Minkowski signature.  Now $\ell^2$ is the determinant of $\ell \cdot \sigma$, so we can re-write the phase space integral as
\be
\int d^4 \ell \delta(\ell_{1 \dot 1} \ell_{2 \dot 2} - \ell_{1 \dot 2} \ell_{2 \dot 1}) M_1(\ell) M_2(-\ell)
\ee
and we can do the integral by, say, integrating over $\ell_{1 \dot 1}$ to give
\be
\frac{d \ell_{2 \dot 2} d \ell_{1 \dot 2} d \ell_{2 \dot 1}}{|\ell_{2 \dot 2}|}
\ee
If we parameterize the remaining integral with $\ell_{a \dot a} = \lambda_{a} \tilde \lambda_{P \dot a}$ and allow $\lambda_{\dot a}$ to run from $-\infty$ to $\infty$, then the integral becomes
\be
\int d^2 \lambda D \tilde \lambda_P M_1(\lambda, \tilde \lambda_P) M_2(\lambda, \tilde \lambda_P)
\ee
where $D \tilde \lambda_P = \langle \tilde \lambda \ d \tilde \lambda \rangle$ is the projective measure on $RP^1$.  It is easy to go from this spinorial representation of the integral to twistor space.  If we Fourier-represent the dependence of $M_1$ and $M_2$ on $\lambda$, we find
\be
\int D \tilde  \lambda_P d^2 \lambda d^2 \tilde \mu_1 d^2 \tilde \mu_2 e^{i [\tilde  \mu_1 - \tilde  \mu_2, \lambda]} M_1(\tilde \lambda, \tilde \mu_1) M_2(\tilde  \lambda, \tilde \mu_2) = \int D^3 W_P M_1(W_P) M_2(W_P)
\ee
with the projective twistor variable $W_P = (\tilde \lambda, \tilde \mu)$.  This is the result we wished to obtain.

We will now make use of the twistor transform in order to represent leading singularities.  A one-loop leading singularity 
\be
\includegraphics[width=9cm]{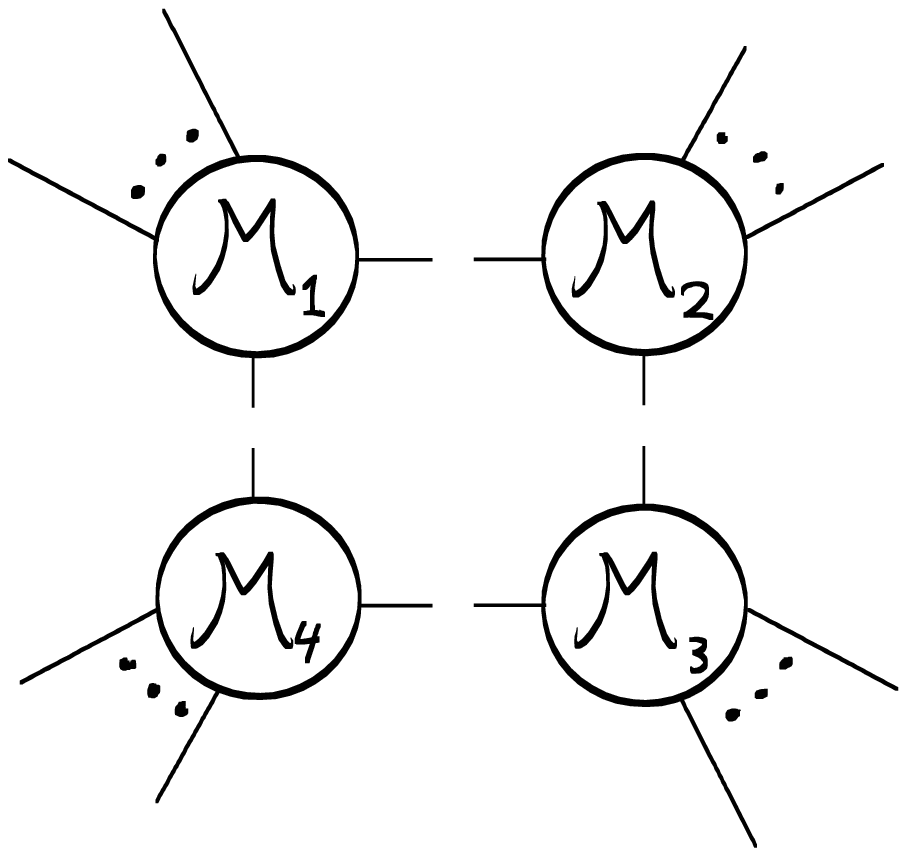} \nonumber
\ee
is given in momentum space by
\be
\int \prod_{i=1}^4 d^4 \ell_i \delta(\ell_i^2) \MM_1(\ell_1,-\ell_2,...) \MM_2(\ell_2,-\ell_3,...) \MM_3(\ell_3,-\ell_4,...) \MM_4(\ell_4,-\ell_1,...) 
\ee
where we are including the momentum conserving delta functions in the tree amplitudes $\MM_i$.  The 16 integration variables in the $\ell_i$ are completely fixed by momentum conservation, which provides 12 constraints, and the condition that $\ell_i^2 = 0$, which provides 4 constraints.  The $\ell_i$ may in general become complex, and we define the integral in this case by analytic continuation\footnote{We will not delve into this issue in detail because it will not be relevant for our analysis, but a more precise definition involves re-interpeting the original loop integral as a contour integral around the four $1/\ell_i^2$ poles}.

In maximally supersymmetric theories we must also sum over the helicities of the particles running in the loop; this is accomplished by integrating over the on-shell superspace variables $\eta$ or $\tilde \eta$ \cite{Nair:1988bq}; for extensive discussions and examples of that formalism see \cite{simplest}.  In twistor space the one-loop leading singularity of ${\cal N}=4$ super Yang-Mills turns into the superconformal integral
\be
\int \prod_{i=1}^4 D^{3|4} \WW_i \MM_1(\WW_1,\WW_2,...) \MM_2(\WW_2,\WW_3,...) \MM_3(\WW_3,\WW_4,...) \MM_4(\WW_4,\WW_1,...) 
\ee
where $\WW = (W, \eta)$.  This has a simple diagrammatic representation as
\be
\includegraphics[width=9cm]{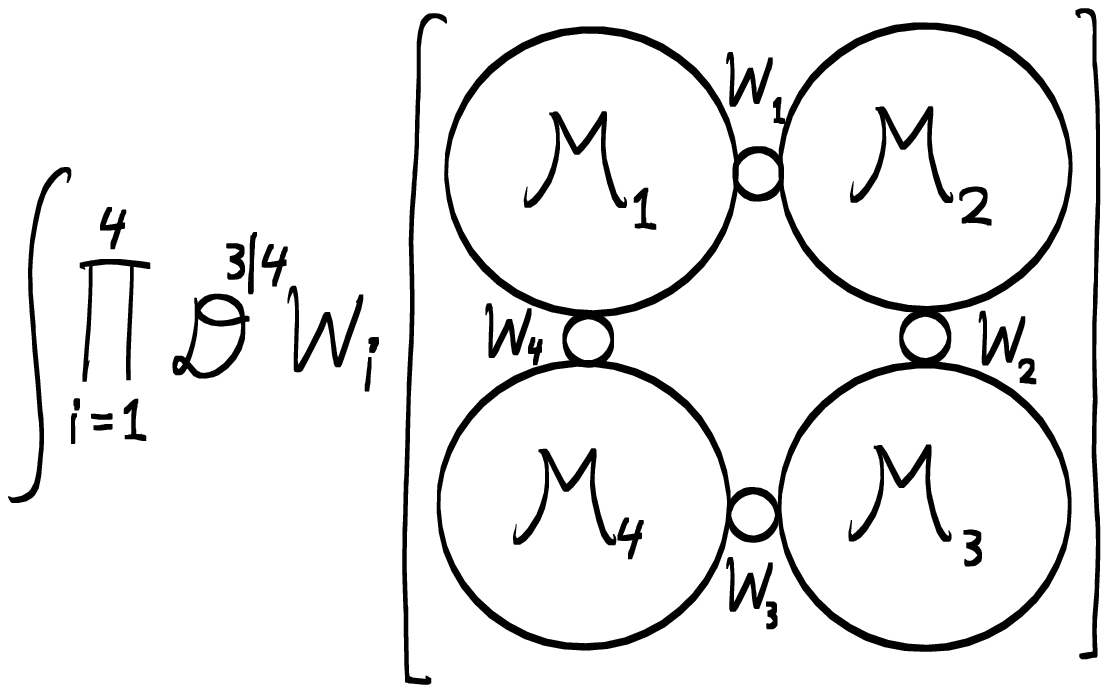} \nonumber
\ee
where we have not explicitly indicated the external states of the amplitudes $\MM_i$.  

It may seem that we have not made much progress, since we have merely substited twistor space integrals for phase space integrals.  However, the twistor space integrands will always be delta functions, so performing the twistor space integrals will only involve some simple linear algebra and book-keeping, making them vastly simpler than their momentum space counterparts.  This fact is an enormous advantage, and it will allow us to begin to unravel the structure of the Grassmannian contour integral ${\cal L}_{n,k}$.

It is straighforward to write higher-loop leading singularities in the same way -- beginning with some $L$ loop diagram with $4L$ propagators, we simply replace each propagator with a twistor variable $\WW_i$, and integrate over it.  Those familiar with `Hodges Diagrams' \cite{Hodges:2005bf}-\cite{ArkaniHamed:2009si}
 may find the picture above familiar, as it is a sort of generalization of those diagrams.  In fact, our diagrammatic representation of leading singularities is in some sense a realization of Hodges' idea of `twistor quilts' \cite{Hodges:2005aj} for loop amplitudes.

\section{Unraveling ${\cal L}_{n,k}$}

In \cite{ArkaniHamed:2009dn} it was conjectured that the residues of the multi-dimensional contour integral
\be
\LL_{n,k}({\cal W}_a) = \int \frac{d^{k \times n} C_{\alpha
a}}{(12\cdots k) \, (23\cdots (k+1) \,) \, \cdots (n 1 \cdots (k-1)
\,)} \prod_{\alpha = 1}^k \delta^{4|4}(C_{\alpha a} {\cal W}_a) 
\ee
are in one-to-one correspondence with the leading singularities of the S-Matrix of ${\cal N}=4$ super Yang-Mills theory.  This conjecture was based in part on evidence accumulated by explicitly computing residues and then identifying them with known leading singularities.  A proof of this conjecture would require a specification of the residues of $\LL_{n,k}$ along with a `dictionary' relating them to the leading singularities.  In this section we will show how any leading singularity can be identified with a residue of $\LL_{n,k}$.

In order to relate leading singularities to residues, we need a way to label them both.  A leading singularity can be specified by drawing an $L$ loop diagram with $4L$ propagators.  When each of these propagators is cut, we will be left with a product of tree amplitudes evaluated with very special kinematics.  If these tree amplitudes are MHV or anti-MHV, then we have a single term, or a `primitive' leading singularity.  Otherwise, we will have a sum of terms, and although one can regard this sum itself as a leading singularity, it is the individual terms in the sum that are residues of $\LL_{n,k}$.  So we should proceed to write each tree amplitude as a sum of terms via the BCFW recursion relations; choosing any one term from each tree amplitude gives a primitive leading singularity.

This last step in the definition may seem a bit arbitrary, but fortunately it can be given a nice interpretation.  As originally shown by Britto, Cachazo, and Feng \cite{bcf}, each term in the BCFW recursion relations can be interpeted as the quadruple cut of a one-loop box (if the tree amplitudes at the corners of the box are all MHV or anti-MHV, this is just a one-loop leading singularity).  This means that wherever we see a non-MHV tree amplitude, we can replace it with a sum over quadruple cuts of one-loop boxes.  This process expresses an $L$ loop object with $4L$ cut propagators in terms of an $L+1$ loop object with $4L + 4$ cut propagators.  If we repeat the process until it terminates, we will be left with a unique product of MHV and anti-MHV tree amplitudes at $L + \delta L$ loops evaluated on the kinematics specified by cutting the $4L + 4 \delta L$ propagators.  Thus each and every term in a leading singularity computed at $L$ loops is in fact itself a leading singularity at $L + \delta L$ loops.

We must also label the residues of $\LL_{n,k}$.  The denominator of $\LL_{n,k}$ is a product of $n$ determinants, so in simple cases it is sufficient to specify on which of these determinants we are evaluating the residue (or in other words, which factors in the denominator vanish).  However, for even moderately large $n$ and $k$ this is inefficient because the residues are highly `composite' \cite{ArkaniHamed:2009dn}, meaning that not only the determinant factors in the denominator vanish, but also their derivatives, second derivatives, and so on.  Furthermore, the equations that follow by requiring that these determinants vanish can have a very large multiplicity of solutions, so this method of labeling does not specify a unique residue.  

This line of thought suggests a better way of labeling the residues.  A residue is given by solving a large system of algebraic equations for coordinates on the Grassmannian, so it is natural to label the residue by the solution itself.  In particular, since points in the Grassmannian can be specified by a $k \times n$ matrix $C_{\alpha a}$ modulo a $GL(k)$ gauge redundancy, it is natural to label residues by specifying the explicit form of $C$.  Naively this sounds like it could be very involved, since one might expect complicated algebraic relationships among the Grassmannian coordinates.  However, we will see that even in very general cases the $C$ matrix takes a form that is both simple and transparently connected to the physics.  For instance, in the case of tree level and one-loop leading singularities we will see that the $C$ matrix can be fully specified by stating which of its entries are zero in a particularly convenient $GL(k)$-gauge.  We will also see that the topology of the loop diagram representing the leading singularity is beautifully reflected by its corresponding Grassmannian locus.

Although we will show how to identify a residue of $\LL_{n,k}$ corresponding to any leading singularity, our analysis will not result in a complete proof that these leading singularities are actually given by the residues in question.  The deficit is due to our inability to compute general composite residues.  This one remaining issue is a precise mathematical problem with a known answer, but its solution should be physically interesting, as the computation of composite residues contains most of the dynamical information of $\LL_{n,k}$.

\subsection{A Simple Tree-Level Illustration}

In \cite{ArkaniHamed:2009si}, \cite{lioneldavid} it was shown that tree level scattering amplitudes in ${\cal N}=4$ super Yang-Mills theory become very simple when transformed to twistor space.  These twistor transformed amplitudes gave way to new expressions for amplitudes in both twistor space and momentum space using the so-called `link representation'.  As an example, the 6-pt NMHV amplitude can be expressed as a sum of terms of the form
\be
U =  \int dc_{iJ} e^{i c_{iJ} W_i \cdot Z_J}  \frac{\delta(c_{52})}{c_{12} c_{32} c_{54} c_{56} c_{14} c_{36} (c_{14} c_{36} - c_{16} c_{34})}  
\ee
in the link representation, where we are ignoring an overall sign factor.  For our purposes, the only thing to notice about this formula is that $c_{52}$ is being set to zero by a delta function.

In \cite{ArkaniHamed:2009dn} we described the contour integral ${\cal L}_{n, k}$, which we conjectured contains all the leading singularities in the ${\cal N}=4$ theory as its residues.  We first discovered this formula by trying to interpret $\delta(c_{52})$ not as a delta function but as a contour integral around the pole $1/c_{52}$.  In fact one can write
\be
{\cal L}_{6, 3} = \int dc_{iJ} e^{i c_{iJ} W_i \cdot Z_J} \frac{1}{c_{52} c_{36} c_{14}(c_{12} c_{54} - c_{14} c_{52})(c_{14} c_{36} - c_{16} c_{34})(c_{36} c_{52} - c_{32} c_{56})}
\ee
and observe that ${\cal L}_{6, 3}$ reduces to the $U$ above on the residue of the pole $c_{52} = 0$.  The discovery of ${\cal L}_{n, k}$ was motivated by a desire to understand how the locality of the S-Matrix is encoded in efficient, on-shell methods such as the BCFW recursion relations, where locality seems to be quite obscure.  In fact as shown in \cite{ArkaniHamed:2009dn} locality is encoded via the very many residue theorems that relate the various residues of ${\cal L}_{n, k}$\footnote{For a different and very interesting approach to this question see \cite{Hodges:2009hk} and also \cite{Korchemsky:2009jv}.}.

However, now that ${\cal L}_{n, k}$ is known, we can reverse the historical logic.  We know that the $U$ above is a term in a 6-pt NMHV tree amplitude, so we could use its explicit form in the link representation to determine which residue of ${\cal L}_{6, 3}$ it comes from.  In what follows we will unravel the embedding of leading singularities among the residues of ${\cal L}_{n, k}$ by identifying them with (very general) link-representation formulas.  In the following three sections we will recursively identify as residues all the one-loop and tree-level leading singularities of $N=4$ super Yang-Mills, and then explain how the method generalizes to arbitrary loop order.  In the appendix we use our method to give an explicit formula for all NMHV ($k=3$) residues.

\subsection{All One-Loop Leading Singularities}

Now we will use what we have learned to identify the residues corresponding to all one-loop leading singularities.  To do this we need only compare the expression for ${\cal L}_{N, K}$ with the integral
\be
\int \prod_{i=1}^4 D^{4|4} \WW_i {\cal L}^i_{n_i,k_i}(\WW_i, \WW_{i+1}, \WW_{a_i})
\ee
This integral can be visualized as the diagram
\be
\includegraphics[width=9cm]{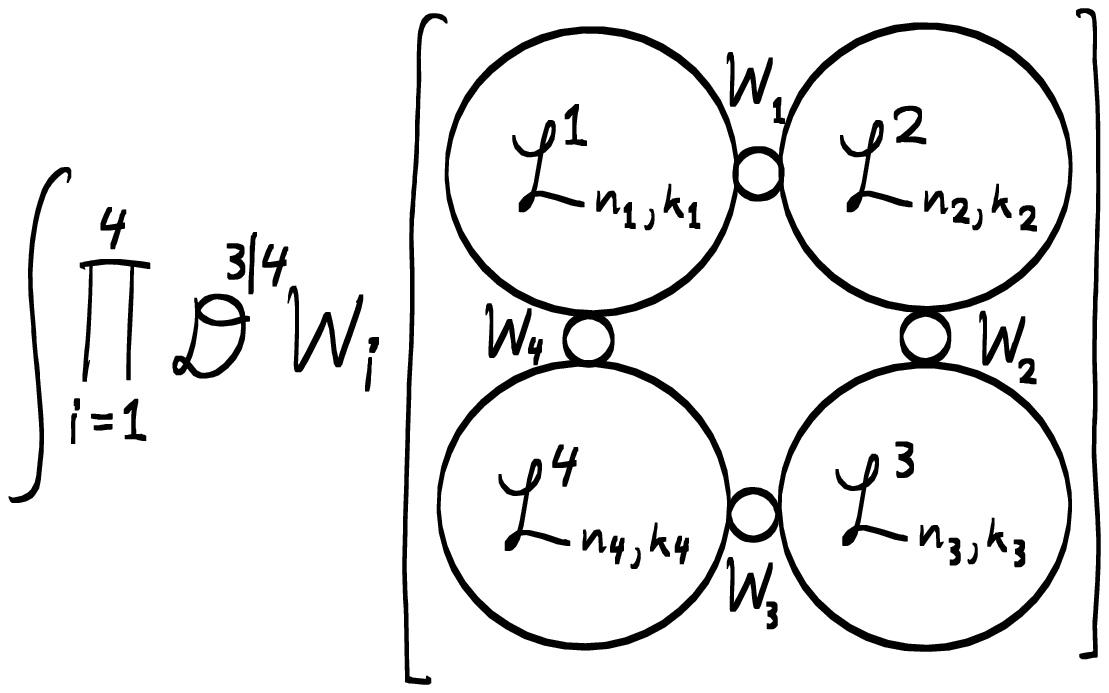} \nonumber
\ee
where we are integrating over the $\WW_i$ with $i=1,2,3,4$, which correspond to the on-shell intermediate propagators in momentum space, and each ${\cal L}^i$ has $n_i-2$ external particles that are not explicitly displayed.  We have labeled the $\WW_{a_i}$ with an index $a_i$ where $i=1,2,3,4$ denotes the particular ${\cal L}_{n_i, k_i}^i$ to which it belongs, and we have a total number of particles $N = n_1 + n_2 + n_3 + n_4 - 8$ and number of negative helicities $K = k_1 + k_2 + k_3 + k_4 - 4$.  Since the four ${\cal L}^i$ depend on the $\WW$ variables only through $\delta^{4|4}(c_{\alpha a} W_a)$, performing the integrals is a matter of book-keeping.

We will choose to only partially fix the $GL(k_i)$ redundancy of the matrices $C^i_{\alpha_i a_i}$ so that
\be
C^i_{\alpha_i a_i} =  \left(\begin{array}{ccccc} 1 & c_{i, 1_i}  & \ldots & c_{i, (n-2)_i} & 0 \\ 0 & C_{\alpha_i, 1_i} & \ldots & C_{\alpha_i, (n-2)_i} & 0 \\ \vdots & \vdots &  \ddots & \vdots & \vdots \\ 0 & C_{\alpha_i, 1_i} & \ldots & C_{\alpha_i, (n-2)_i} & 0 \\ 0 & c_{i+1, 1_i} & \ldots & c_{i+1, (n-2)_i} & 1 \end{array} \right)
\ee
or in other words, we have fixed the first and last columns of the matrix, which correspond to the $\WW_i$ variables over which we are going to integrate, but we have not fixed the other columns and rows.  There is a subtlety when treating the anti-MHV 3-pt amplitude, because its $C$ matrix has only a single row and therefore we can only fix it to be $C = (1, c_{1 1_i}, c_{12})$.  We will return to treat this special case at the end.

We will choose to use a delta function from ${\cal L}^i$ to perform the integral over $\WW_i$.  Naively one would expect to simply solve for the $\WW_i$, but the twistor variables are projective, so we can only conclude that 
\be
\WW_i = \tau_i \sum_{a_i} -c_{i, a_i} \WW_{a_i} 
\ee
for some non-zero $\tau_i$.  This new $\tau_i$ is an arbitrary parameter, so we can use it to fix one of the $c$ variables, so e.g. we could set $c_{i,1_i} = 1$ for each $i=1,2,3,4$.  This follows because we can then absorb $\tau_i$ everywhere else it appears by re-scaling the other variables.  However we will ignore the $\tau_i$ for now in order to avoid breaking any symmetries.  $\WW_i$ appears in both ${\cal L}^i$ and ${\cal L}^{i-1}$; subsituting it into the latter takes
\be
\delta^{4|4}(c_{i, a_{i-1}} \WW_{a_{i-1}} + \WW_i) \to \delta^{4|4}(c_{i, a_{i-1}} \WW_{a_{i-1}} - c_{i, a_i} \WW_{a_i})
\ee
and now we are done!  The one-loop leading singularity corresponds to ${\cal L}_{N, K}$ with $C$ matrix fixed to the form
\be
\label{OneLoopLS}
C =  \left(\begin{array}{cccccccccccc} c_{1, 1_1} & \ldots & c_{1, (n - 2)_1} & 0 & \ldots & 0 & 0 & \ldots & 0 & c_{\alpha, 1_4} & \ldots & c_{\alpha, (n-2)_4} \\ 
c_{\alpha, 1_1} & \ldots & c_{\alpha, (n - 2)_1} & 0 & \ldots & 0 & 0 & \ldots & 0 & 0 & \ldots & 0 \\ 
\vdots &  \vdots & \vdots &  \vdots & \vdots & \vdots & \vdots & \vdots & \vdots & \vdots & \vdots & \vdots\\
c_{\alpha, 1_1} & \ldots & c_{\alpha, (n - 2)_1} & 0 & \ldots & 0 & 0 & \ldots & 0 & 0 & \ldots & 0 \\ 
 c_{\alpha, 1_1} & \ldots & c_{\alpha, (n - 2)_1} & c_{2, 1_2} & \ldots & c_{2, (n-2)_2} & 0 & \ldots & 0 & 0 & \ldots & 0 \\ 
0 & \ldots & 0 & c_{\alpha, 1_2} & \ldots & c_{\alpha, (n-2)_2} & 0 & \ldots & 0 & 0 & \ldots & 0 \\ 
\vdots &  \vdots & \vdots &  \vdots & \vdots & \vdots & \vdots & \vdots & \vdots & \vdots & \vdots & \vdots\\
0 & \ldots & 0 & c_{\alpha, 1_2} & \ldots & c_{\alpha, (n-2)_2} & c_{3, 1_3} & \ldots & c_{3, (n-2)_3} & 0 & \ldots & 0 \\ 
0 & \ldots & 0 & 0 & \ldots & 0 & c_{\alpha, 1_3} & \ldots & c_{\alpha, (n-2)_3} & 0 & \ldots & 0 \\ 
\vdots &  \vdots & \vdots &  \vdots & \vdots & \vdots & \vdots & \vdots & \vdots & \vdots & \vdots & \vdots\\
0 & \ldots & 0 & 0 & \ldots & 0 & c_{\alpha, 1_3} & \ldots & c_{\alpha, (n-2)_3} & 0 & \ldots & 0 \\ 
0 & \ldots & 0 & 0 & \ldots & 0 & c_{\alpha, 1_3} & \ldots & c_{\alpha, (n-2)_3} & c_{4, 1_4} & \ldots & c_{4, (n-2)_4} \\ 
0 & \ldots & 0 & 0 & \ldots & 0 & 0 & \ldots & 0 & c_{\alpha, 1_4} & \ldots & c_{\alpha, (n-2)_4}  \\ 
\vdots &  \vdots & \vdots &  \vdots & \vdots & \vdots & \vdots & \vdots & \vdots & \vdots & \vdots & \vdots\\
0 & \ldots & 0 & 0 & \ldots & 0 & 0 & \ldots & 0 & c_{\alpha, 1_4} & \ldots & c_{\alpha, (n-2)_4}  \\ 
\end{array} \right) 
\ee
where we have eliminated the minus signs in front of the $c_{i, a_i}$ variables by a simple redefinition.  We did not completely fix the $GL(k_i)$ gauge redundancies of the ${\cal L}^i_{n_i, k_i}$ in order to avoid obscuring the structure of this matrix, but in practical computations one would fix these redundancies in some way.  Also, although we have written the matrix as almost-block-diagonal, the diagonal of the matrix plays no special role -- we are free to cyclicly permute the columns and rows.  We should think of this $C$ matrix `picture' as a specification of the linear dependencies among its  various columns.

Let us count the number of free variables in momentum space to show that the contour of integration has been completely specified.  After fixing the $GL(k_i)$ redundancies and choosing a particular residue for the ${\cal L}_{n_i, k_i}^i$ we are left with $2 n_i - 4$ variables in each ${\cal L}^i$ \cite{ArkaniHamed:2009dn}, which would be fixed by delta functions were we to transform back to momentum space.  This means that there are a total of $2N$ free variables after the individual ${\cal L}^i$ contours have been specified.  However, we saw above that there are four $\tau_i$ parameters which appear as a consequence of the fact that we have integrated over $R^4$ instead of $RP^3$ four times;  we can use these to eliminate four $c$ variables by setting them to $1$.  If we take ${\cal L}_{N, K}$ to momentum space we find $2N-4$ delta function constraints, which is exactly equal to the number of free variables.

Before giving some examples let us return to the case where one of the ${\cal L}^i$, say ${\cal L}^1$, is an anti-MHV 3-pt amplitude. Let us fix its $C$ `matrix' to be
\be
C = (1, c_{1 1_1}, c_{1 2})
\ee
so that the amplitude becomes
\be
{\cal L}^1_{3,1} = \int \frac{d c_{1 1_1} dc_{1 2}}{c_{1 1_1} c_{1 2}} \delta^{4|4}(\WW_1 + c_{1 1_1} \WW_{1_1} + c_{1 2} \WW_2)
\ee
In accord with our choices above we will use this delta function to integrate over $\WW_1$, giving
\be
\WW_1 = \tau_1 (-c_{1 1_1} \WW_{1_1} - c_{1 2} \WW_2) \to c_{1 1_1} \WW_{1_1} +  \WW_2
\ee
with an appropriate choice of the free parameter $\tau_1$ and re-scaling of $c_{1 1_1}$.  Now we have completely eliminated ${\cal L}^1$, its only remnant being the $c_{1 1_1}$ parameter.  As before, we will solve for $\WW_2$ using a delta function from ${\cal L}^2$, so the end result is a $C$ matrix for ${\cal L}_{N,K}$ of the form
\be
C =  \left(\begin{array}{cccccccccccc} c_{1, 1_1} & c_{2, 1_2} & \ldots & c_{2, (n-2)_2} & 0 & \ldots & 0 & c_{\alpha, 1_4} & \ldots & c_{\alpha, (n-2)_4} \\ 
0 & c_{\alpha, 1_2} & \ldots & c_{\alpha, (n-2)_2} & 0 & \ldots & 0 & 0 & \ldots & 0 \\ 
\vdots &  \vdots & \vdots & \vdots & \vdots & \vdots & \vdots & \vdots & \vdots & \vdots\\
0 & c_{\alpha, 1_2} & \ldots & c_{\alpha, (n-2)_2} & 0 & \ldots & 0 & 0 & \ldots & 0 \\ 
0 & c_{\alpha, 1_2} & \ldots & c_{\alpha, (n-2)_2} & c_{3, 1_3} & \ldots & c_{3, (n-2)_3} & 0 & \ldots & 0 \\ 
0 & 0 & \ldots & 0 & c_{\alpha, 1_3} & \ldots & c_{\alpha, (n-2)_3} & 0 & \ldots & 0 \\ 
\vdots &  \vdots & \vdots & \vdots & \vdots & \vdots & \vdots & \vdots & \vdots & \vdots\\
0 & 0 & \ldots & 0 & c_{\alpha, 1_3} & \ldots & c_{\alpha, (n-2)_3} & 0 & \ldots & 0 \\ 
0 & 0 & \ldots & 0 & c_{\alpha, 1_3} & \ldots & c_{\alpha, (n-2)_3} & c_{4, 1_4} & \ldots & c_{4, (n-2)_4} \\ 
0 & 0 & \ldots & 0 & 0 & \ldots & 0 & c_{\alpha, 1_4} & \ldots & c_{\alpha, (n-2)_4}  \\ 
\vdots &  \vdots & \vdots & \vdots & \vdots & \vdots & \vdots & \vdots & \vdots & \vdots\\
0 & 0 & \ldots & 0 & 0 & \ldots & 0 & c_{\alpha, 1_4} & \ldots & c_{\alpha, (n-2)_4}  \\ 
\end{array} \right) 
\ee

Let us now check these very general results with a few examples.  If we want to obtain a box coefficient (one-loop leading singularity) in the MHV sector, we must make one pair of opposite corners MHV and the other pair anti-MHV 3-pt amplitudes.  This gives a $C$ matrix structure
\be
C =  \left(\begin{array}{cccccccc} 
* & * & \ldots & * & 0 & * \cdots & *\\ 
0 & * & \ldots & * & * & * \cdots & *\\ 
\end{array} \right) 
\ee
where there are still two $\tau$ parameters to be specified (in other words, we can rescale the two rows independently by an arbitrary factor, setting a $c$ parameter in each equal to $1$).  

One might wonder what would have happened if we made the two anti-MHV 3-pt amplitudes adjacent.  Physically, this sort of leading singularity must vanish; our results give 
\be
C =  \left(\begin{array}{ccccc} 
* & * & * & \cdots & *\\ 
0 & 0 & * & \cdots & *\\ 
\end{array} \right) 
\ee
In this case the sub-determinant $(1,2)$ vanishes.  If we interpret this as $1/0$ it means that our result is not well-defined.  If we attempt to view ${\cal L}_{n,2}$ as a contour integral evaluated on the residue $(1,2)$, then when we return to momentum space we would find an additional constraint on the momenta beyond momentum conservation, or in other words we would find that this object vanishes for generic momenta.  Thus we see that ${\cal L}_{n,k}$ ``knows'' that this is not a viable leading singularity.

Finally let us consider a much more non-trivial example.  In the case $N=8$, $K=4$ there is a single four mass box which corresponds to $n_i=4$, $k_i=2$ for all $i$, or in other words this is a box with a 4-pt MHV amplitude at each corner.  Eliminating the four extra variables, we obtain a matrix structure
\be
C = \left(\begin{array}{cccccccc} * & 1 & 0 & 0 & 0 & 0 & * & * \\ 
* & * & * & 1 & 0 & 0 & 0 & 0 \\
0 & 0 & * & * & * & 1 & 0 & 0 \\
0 & 0 & 0 & 0 & * & * & * & 1 
\end{array} \right) 
\ee
We immediately see that the determinants $(I, I+1, I+2, I+3)$ vanish for $I$ odd but that they are non-vanishing for $I$ even.  This was precisely the residue found in \cite{ArkaniHamed:2009dn} to correspond to this particular leading singularity.

\subsubsection*{A Worked Example}

In the analysis above we saw how one-loop leading singularities correspond to particular $C$ matrix structures, or in other words, to particular subspaces of the Grassmannian.  However, we did not show how one obtains these $C$ matrices from contour integration, and we did not work out the resulting residues.  We will go through these procedures in detail for the $n=8$, $k=4$ example, and then we will explain how they generalize.

We would like to fix the $GL(4)$ redundandancy so that 
\be
C = \left(\begin{array}{cccccccc} c_{21} & 1 & x_3 & 0 & 0 & 0 & c_{27} & c_{28} \\ 
c_{41} & c_{42} & c_{43} & 1 & x_5 & 0 & 0 & 0 \\
0 & 0 & c_{63} & c_{64} & c_{65} & 1 & x_7 & 0 \\
x_1 & 0 & 0 & 0 & c_{85} & c_{86} & c_{87} & 1 
\end{array} \right) 
\ee
However, there is a non-trivial Jacobian that arises when we fix the $GL(4)$ redundancy in this way.  The easiest way to compute this Jacobian is to write our $C$ matrix as a $GL(4)$ transformation acting on an `old' matrix
\be
C^{\mathrm{old}} = \left(\begin{array}{cccccccc} 
* & 1 & * & 0 & * & 0 & * & 0  \\ 
* & 0 & * & 1 & * & 0 & * & 0  \\
* & 0 & * & 0 & * & 1 & * & 0 \\
* & 0 & * & 0 & * & 0 & * & 1 
\end{array} \right)  
\ee
so that
\be
C^{\mathrm{new}}_{\alpha a} = J_{\alpha}^{\ \beta}(C^{\mathrm{new}})  C^{\mathrm{old}}_{\beta a}
\ee
We know that the measure is simply $d^{k(n-k)}C^{\mathrm{old}}$, so we can compute the Jacobian in terms of the new variables using $J^{-1} C^{\mathrm{new}}$.  It is straightforward to compute this Jacobian in general, which we have done in the appendix.  In our case, the Jacobian is
\be
(c_{41} - c_{21} c_{42}) (c_{63} - c_{43} c_{64}) (c_{85} - c_{65} c_{86}) (c_{27} - c_{87} c_{28})
\ee
The product of $4 \times 4$ determinants in the denominator of the integrand of ${\cal L}_{8,4}$ is
\be
\prod_{i=1}^8 D_i  & = & (x_1(c_{63} - c_{43} c_{64} + x_3 c_{42} c_{64})) (c_{85}(c_{63} - c_{43} c_{64} + x_3 c_{42} c_{64})) ... \nonumber \\
& = &  (c_{41} - c_{21} c_{42})^2 (c_{63} - c_{43} c_{64})^2 (c_{85} - c_{65} c_{86})^2 (c_{27} - c_{87} c_{28})^2 \nonumber \\ 
& & \times  c_{41} c_{63} c_{85} c_{27} \cdot x_1 x_3 x_5 x_7 \ + \ O(x^5)
\ee
Note that the four factors on the first line are squared, but one of each will be canceled by the Jacobian.  Taking this into account, we see that ${\cal L}_{8,4}$ takes the simple form
\be
{\cal L}_{8,4} & = & \oint \frac{dx_1 dx_3 dx_5 dx_7}{x_1 x_3 x_5 x_7} \nonumber \\
& & \times \int \frac{d^{12} c_{i J} \ \delta^{4|4}( C_{\alpha a} \WW_a)}{c_{41} c_{63} c_{85} c_{27} (c_{41} - c_{21} c_{42}) (c_{63} - c_{43} c_{64}) (c_{85} - c_{65} c_{86}) (c_{27} - c_{87} c_{28})}
\ee
The contour integral over the $x$ immediately sets them all to zero, so we have neglected higher order terms in these variables.  The denominator is precisely what we get from the denominators of the four MHV amplitudes at the corners of the one-loop leading singularity (i.e. the `box coefficient'; note that four $c$ parameters have been eliminated using $\tau$ variables).

Now we can Fourier transform from twistor space back to momentum space.  The most general way to do this is to write
\be
\int d^2 \mu_a e^{i [\tilde \lambda_a \mu_a]} \delta^{4}(C_{\alpha a} W_a) = \delta^{2}(C_{\alpha a} \tilde \lambda_a) \int d^2 \rho_{\alpha} \delta^2(\lambda_a - C_{\alpha a} \rho_\alpha)
\ee
so now the $c$ variables must satisfy 
\be
C_{\alpha a} \tilde \lambda_a = 0 \ \ \ \mathrm{and} \ \ \ \lambda_a - C_{\alpha a} \rho_\alpha = 0
\ee
where the $\rho_\alpha$ are auxiliary spinor variables.  Clearly the first set of equations is linear in the $C$ variables.  However, because the auxiliary $\rho_\alpha$ are free, the second set of equations is in general quadratic.  Something interesting has occurred, as the entire kinematic structure of the leading singularity is encoded in these simple quadratic equations!  Note also that any multiplicity of solutions will come entirely from these momentum space equations.  We expect that in general the multiplicity will exactly match the multiplicity of solutions to the $4L$ cut conditions at $L$ loops.

The procedure that we have followed generalizes to the computation of any one-loop leading singularity, with one crucial caveat -- in general, the contour integral over the $x$ variables will not be so simple.  We will generically have a large $C$ matrix, the number of $x$ variables will be much larger than the number of external particles, and the residue at $x=0$ will be highly composite.    However, we have a very definite expectation, namely that this residue must equal the product of the four $\LL_{n_i, k_i}$ denominators.  In the appendix we argue for the existence of the residue, but we do not know how to compute it and prove that our expectation is correct.

\subsection{Back to BCF}

\begin{figure}[htp]
\begin{center}
\includegraphics[width=14cm]{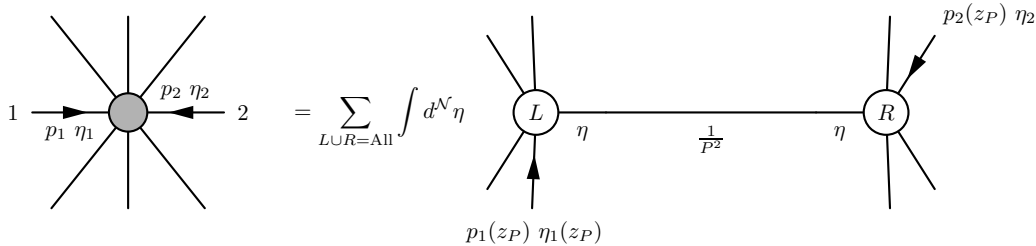}
\caption{The BCFW Recursion Relations in maximally supersymmetric theories.}
\end{center}
\end{figure}

The BCFW recursion relations \cite{bcf}-\cite{Drummond:2008cr} are an extremely efficient method for computing tree level scattering amplitudes in a variety of theories.  Some key features of these recursion relations are that they compute scattering amplitudes using purely on-shell information, and that they assemble local amplitudes from non-local pieces.  As an example, the 6-pt amplitude in Yang-Mills theory is
\be
M^{+-+-+-}_{{\rm BCFW}} & = & \left( 1 + r^2 + r^4 \right) \,\left[\frac{\langle 4 6 \rangle^4 [1 3]^4}{[12][23]\langle 45 \rangle \langle 56 \rangle (p_4 + p_5 + p_6)^2} \right.
\nonumber \\ & & \left. \times \frac{1}{\langle 6 |5 + 4|3] \langle 4|5 + 6 |1]} \right]
\ee
when computed with BCFW (where $r$ cyclicly permutes the external particles $i \to i+1$).  Note that the factor on the second line has unphysical poles, and therefore it could never come from the Feynman diagrams of a local theory.  One of the main motivations underlying the discovery of ${\cal L}_{n,k}$ was to find a way to explain how local amplitudes arise from non-local pieces.

The BCFW recursion relations were originally discovered by Britto, Cachazo, and Feng \cite{bcf} in a study of the IR equations as applied to one-loop leading singularities \cite{Roiban:2004ix}-\cite{lead}.  This means that each term in the recursion relations is a one-loop leading singularity, so we can use our techniques from the previous subsection to identify the contours of integration in ${\cal L}_{n,k}$ that correspond to tree amplitudes.

To be more specific, we want to look at one-loop leading singularities with $(n_1,k_1) = (3,2)$ and $(n_2,k_2)=(3,1)$, or in other words we take these two neighboring corners of the box to be MHV and anti-MHV 3-pt amplitudes  
\be
\includegraphics[width=13cm]{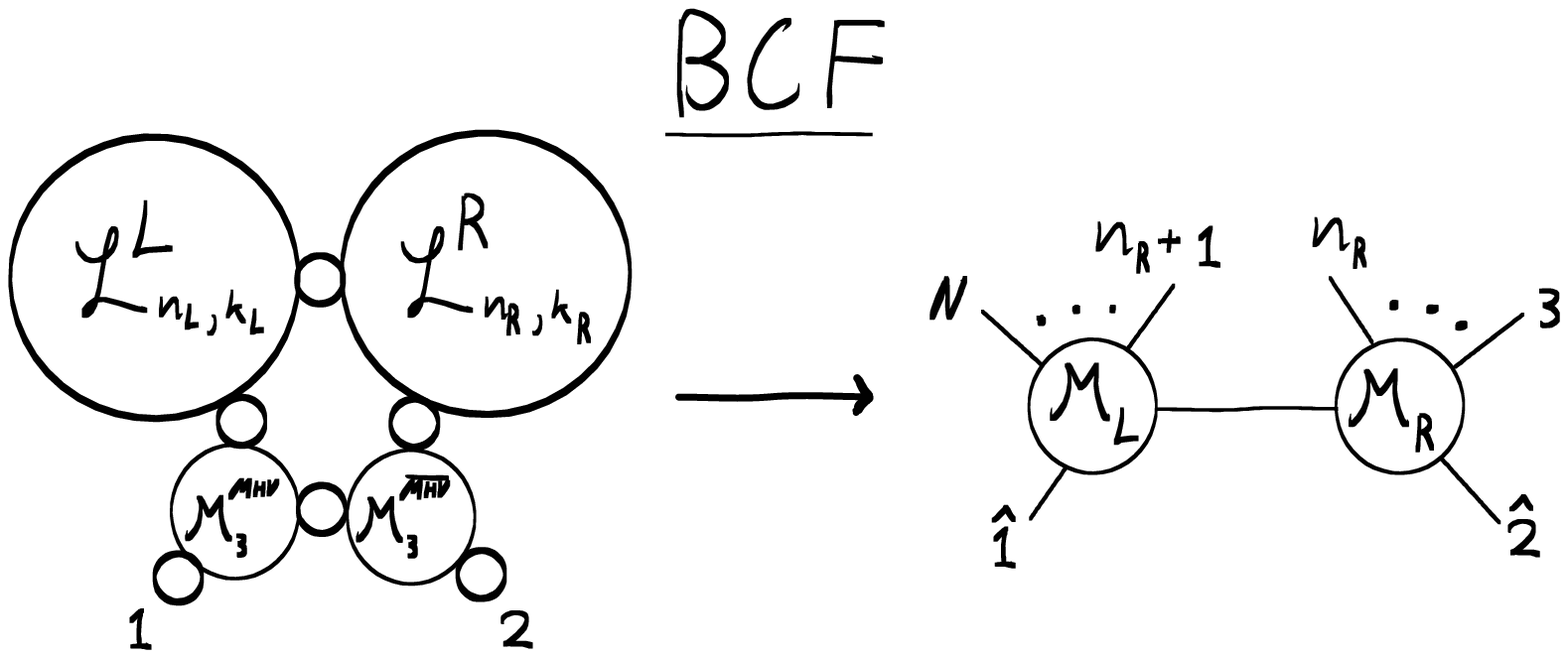} \nonumber
\ee
where particles $1$ and $2$ correspond to the analytically continued particles in the BCFW recursion relations, and the small unlabeled circles represent projective $\WW$ variables to be integrated over.  Using our solution from the previous subsection, we find a $C$ matrix in ${\cal L}_{N,K}$ of the form
\be
C =  \left(\begin{array}{cccccccc} 
c_{1 1_1} & 1 & 0 & \ldots & 0 & c_{\alpha, 2_L} & \ldots & c_{\alpha, (n-1)_L} \\ 
0 & c_{2 1_2} & c_{2, 2_R} & \ldots & c_{2, (n-1)_R} & 0 & \ldots & 0 \\ 
0 & 0 & c_{\alpha, 2_R} & \ldots & c_{\alpha, (n-1)_R} & 0 & \ldots & 0 \\ 
\vdots & \vdots & \vdots & \vdots & \vdots & \vdots & \vdots & \vdots\\
0 & 0 & c_{\alpha, 2_R} & \ldots & c_{\alpha, (n-1)_R} & 0 & \ldots & 0 \\ 
0 & 0 & c_{I, 2_R} & \ldots & c_{I, (n-1)_R} & c_{I, 2_L} & \ldots & c_{I, (n-1)_L} \\ 
0 & 0 & 0 & \ldots & 0 & c_{\alpha, 2_L} & \ldots & c_{\alpha, (n-1)_L}  \\ 
\vdots & \vdots & \vdots & \vdots & \vdots & \vdots & \vdots & \vdots\\
0 & 0 & 0 & \ldots & 0 & c_{\alpha, 2_L} & \ldots & c_{\alpha, (n-1)_L}  \\ 
\end{array} \right) 
\ee
where we have indexed most of the $c$'s with $L$ and $R$ to show that these belong to the usual $\MM_L$ and $\MM_R$ of BCFW, and we have used a label $I$ for `intermediate' for the one overlapping row.   It is worth noting that this matrix structure is not so surprising -- it is perhaps the first thing one might guess.  The BCFW form of the amplitude is being represented by two blocks that correspond to $\MM_L$ and $\MM_R$ and which share a row that corresponds to the intermediate particle.

Let us check our general formula with a few examples.  The simplest example is the computation of an MHV amplitude by BCFW; for this case we would find a $C$ matrix
\be
C =  \left(\begin{array}{cccccc} 
* & 1 & 0 & * & \cdots & *\\ 
0 & * & * & * & \cdots & *\\ 
\end{array} \right) 
\ee
None of the sub-determinants $(I, I+1)$ vanish, which is exactly what we would expect for the $C$ matrix of an MHV amplitude.  The $GL(2)$ symmetry has not been fully fixed because we have yet to use the $\tau_L$ projectivity parameter, we can use it to obtain the fully fixed matrix
\be
C =  \left(\begin{array}{cccccc} 
* & 1 & 0 & * & \cdots & *\\
0 & * & 1 & * & \cdots & *\\
\end{array} \right) 
\ee
from which one could compute the MHV amplitude in momentum space.

As another example, consider the 6-pt NMHV amplitude.  One of the terms used to construct it comes from applying BCFW where $\MM_L$ and $\MM_R$ are both 4-pt amplitudes.  In this case we would take the $C$ matrix to be
\be
C =  \left(\begin{array}{cccccc} 
* & 1 & 0 & 0 & * & *\\ 
0 & * & * & * & 0 & 0\\ 
0 & 0 & * & * & * & *\\ 
\end{array} \right) 
\ee
We see that only one of the determinants $(I, I+1, I+2)$ vanishes, namely the one with $I = 5$.  This is precisely what was found in \cite{ArkaniHamed:2009dn}.  As another example, consider again the 6-pt amplitude constructed from a 5-pt and a 3-pt MHV amplitude, this would have $C$ matrix
\be
C =  \left(\begin{array}{cccccc} 
* & 1 & 0 & 0 & 0 & *\\ 
0 & * & * & * & * & 0\\ 
0 & 0 & * & * & * & *\\ 
\end{array} \right) 
\ee
so we see that the $I=3$ determinant vanishes.

Using these results one can recursively identify the contours of integration that correspond to tree amplitudes.  A $C$ matrix of the form that we have identified in this section will give terms that can contribute to tree level amplitudes as long as the contours of integration for $c_L$ and $c_R$ are chosen to give components of tree level amplitdues.  To obtain the full BCFW recursion relations one simply sums over the sets $L$ and $R$ with appropriate contours for the sub-Grassmannians.

The analysis of this subsection and the last is one-half of a constructive proof that all one-loop leading singularities are contained in ${\cal L}_{n,k}$ for some choice of contour.  This follows because (1) we have (recursively) shown that all tree amplitudes are contour integrals via BCF and the IR equations, and (2) we have identified the contour for one-loop leading singularities given the sub-contours for the four tree amplitudes at the corners of the box.  Our method also partially explains why terms that come from non-adjacent BCFW deformations do not arise as residues of ${\cal L}_{n,k}$ -- due to color ordering, these terms cannot be written as one-loop leading singularities and so they cannot be found among the residues.  However, our results are only half of a proof because we have not explicitly computed the residues themselves, but only shown how to obtain the appropriate contours of integration.

\subsection{Higher Loops and General Patterns}

\begin{figure}[htp]
\begin{center}
\includegraphics[width=16cm]{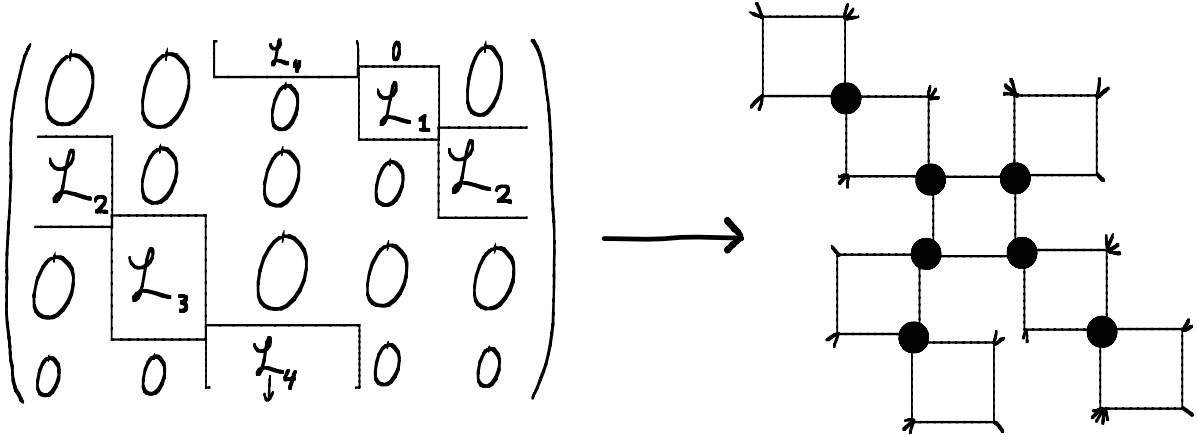} \nonumber
\caption{A diagram representing the infinite class of leading singularities that can be identified by applying our one-loop analysis recursively, expanding the objects at the corners of a one-loop box into new boxes.  The heavy black dots are four point amplitudes linking boxes together at their corners.}\label{BoxWeb}
\end{center}
\end{figure}
In the previous sections we identified the contours of integration for all tree amplitudes and all one-loop leading singularities.  These results immediately apply to an infinite class of leading singularities -- those that can be constructed by attaching `boxes' together at their corners.  This follows because we can recursively interpret each of the 4 sub-matrices of ${\cal L}_{N, K}$ as one-loop leading singularities themselves.  Thus the general statement is that this type of leading singularity at $L$ loops corresponds to a configuration where ${\cal L}_{N, K}$ is broken up into $3L+1$ submatrices following the pattern of equation (\ref{OneLoopLS}).  This structure of leading singularity can be visualized as in figure \ref{BoxWeb} where the black dots represent the joined corners and the little tick marks at the other corners represent external particles. 

These sorts of leading singularities always correspond to block diagonal $C$ matrices (note that the fact that the blocks lie on the diagonal is itself meaningless because we are free to cyclicly translate all of the columns).  This makes sense based on the topological structure of the loop diagram, because beginning at any point on the diagram one can follow propagators and ``walk'' from tree amplitude to tree amplitude, encountering every propagator and tree amplitude in cyclic order.  For more general topologies this would not be possible -- one would inevitably miss some tree amplitudes and propagators.

\begin{figure}[htp]
\begin{center}
\includegraphics[width=12cm]{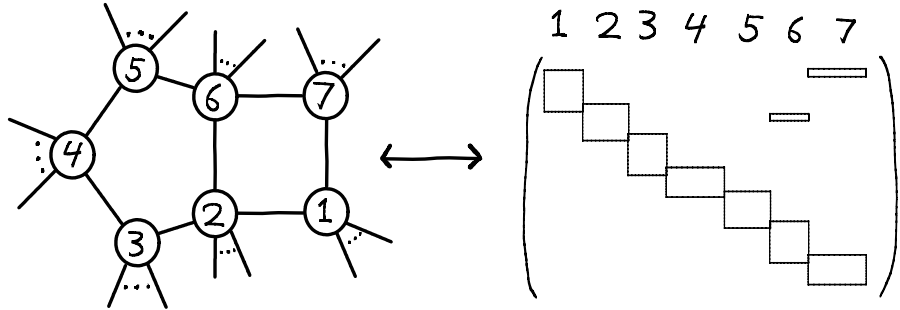}
\caption{This diagram shows a 2-loop leading singularity and the corresponding points in the Grassmannian to which it corresponds.  The rectangles in the pictured matrix correspond to its non-zero entries, and the adjacent boxes share a single row.}\label{BoxPentagon}
\end{center}
\end{figure}

We can write more general leading singularities in twistor space using the method of section 2.  As a first example we can consider the diagram of figure \ref{BoxPentagon}.  Computing this diagram in twistor space is straightforward, since again we only need to integrate over delta functions.  We will not go through the computation in detail or consider the possible subtleties that can arise when the various tree amplitudes at the corners have too few delta functions (i.e. for very small $n_i$ and $k_i$). We will only give the generic result because our goal is to explicate the pattern of how leading singularities correspond to various sub-structures in the $k,n$ Grassmannian.  

The easiest way to compute this leading singularity is to first take account of the propagators around the borders of the box and pentagon and only then integrate over the single $\WW$ variable corresponding to the propagator shared between the box and the pentagon.  The first step gives a structure in ${\cal L}_{N,K}$ that is block diagonal as in the one-loop case except with $7$ blocks instead of $4$.  The second step eliminates a row and column, with the result that two non-adjacent blocks now share a row.  This can be pictured as in figure
\ref{BoxPentagon}, where we have explicitly displayed the $C$ matrix structure that arises when this leading singularity is embedded in ${\cal L}_{N,K}$ (the regions outside the boxes are filled with zeroes).  This analysis can be generalized to another infinite class of leading singularities made up of boxes and pentagons that are chained together along various sides in such a way that there are $4L$ propagators at $L$ loops.  

Although we have given the general case above, we have also explicitly checked our results for the case $n=12$, $k=6$, ie for an N$^4$MHV amplitude.  In that case the full $C$ matrix for the 2-loop leading singularity
\be
\includegraphics[width=6cm]{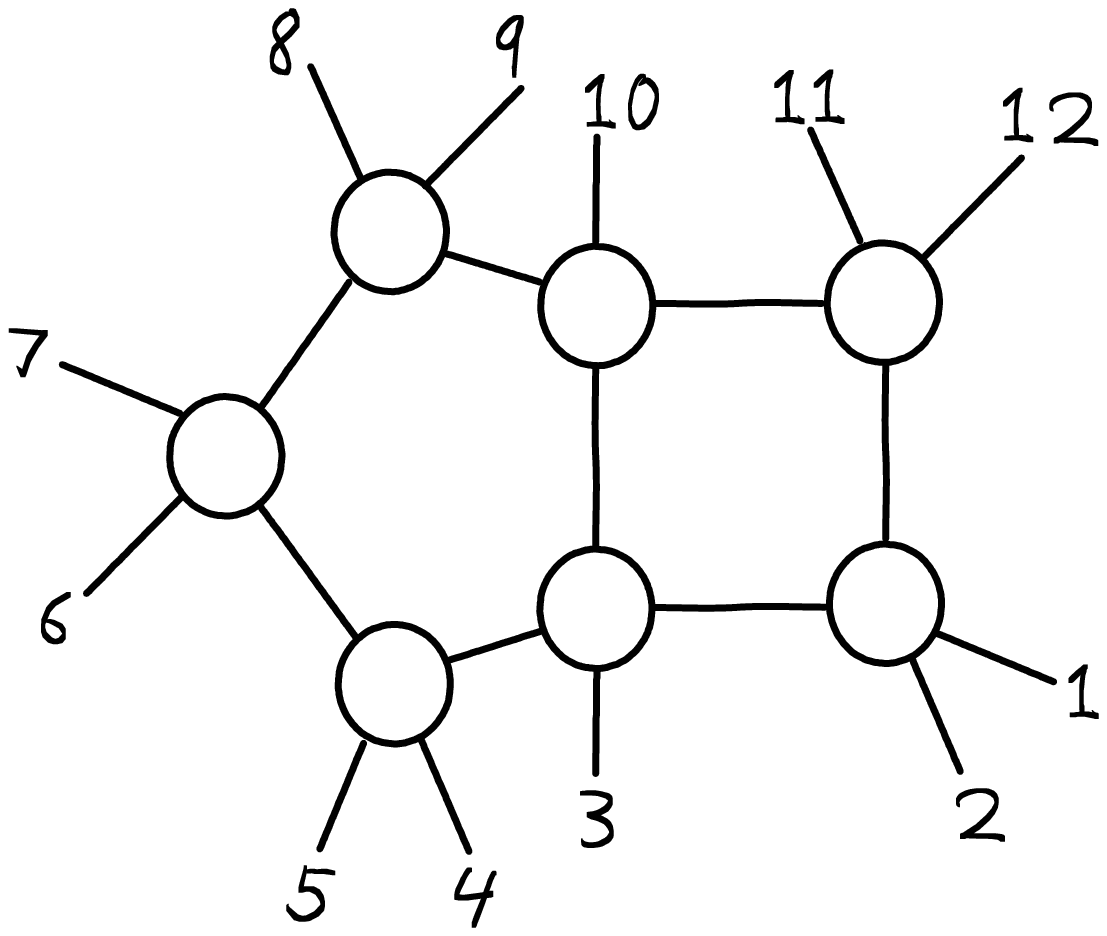} \nonumber
\ee
takes the form
\be
C = \left(\begin{array}{cccccccccccccc} 
c_{1,1} & 1 & 0 & 0 & 0 & 0 & 0 & 0 & 0 & 0 & c_{1,11} & c_{1,12} \\
a & b & c_{2,3} & c_{2,4} & c_{2,5} & 0 & 0 & 0 & 0 & 0 & 0 & 0 \\
0 & 0 & 0 & c_{3,4} & c_{3,5} & c_{3,6} & 1 & 0 & 0 & 0 & 0 & 0 \\
0 & 0 & 0 & 0 & 0 & c_{4,6} & c_{4,7} & c_{4,8} & 1 & 0 & 0 & 0 \\
a & b & 1 & 0 & 0 & 0 & 0 & c_{5,8} & c_{5,9} & c_{5,10} & 0 & 0 \\
a & b & 1 & 0 & 0 & 0 & 0 & 0 & 0 & c_{6,10} & c_{6,11} & c_{6,12}
\end{array} \right) 
\ee
This two-loop leading singularity has a very complicated kinematic structure in momentum space.  By this we mean that when one solves the $8$ quadratic equations that force the $8$ intermediate propagators on-shell, the solution involves elaborate double square roots of kinematic invariants.  When ${\cal L}_{12,6}$ is transformed to momentum space one obtains the equations
\be
C_{\alpha a} \tilde \lambda_a = 0 \ \ \ \mathrm{and} \ \ \ \lambda_a - C_{\alpha a} \rho_\alpha = 0
\ee
for the $c_{iJ}$ and $a$ and $b$ variables, where $\rho_\alpha$ are auxiliary spinors that must be solved for and eliminated.  We have checked explicitly\footnote{with the help of Jacob Bourjaily} that with our $C$ matrix structure these equations give precisely the kinematic structure of the leading singularity.  This is an extremely non-trivial check of our methods and of the claim that the residues of $\LL_{n,k}$ are in fact leading singularities.

\begin{figure}[htp]
\begin{center}
\includegraphics[width=14cm]{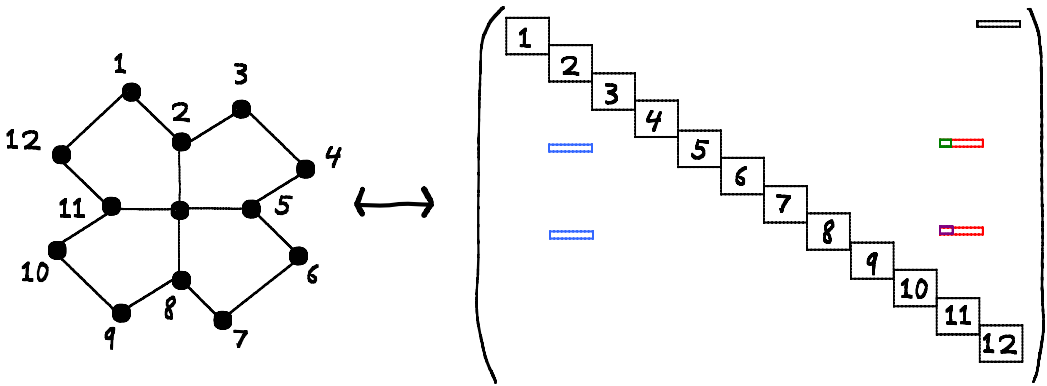}
\caption{An example of a 4-loop leading singularity and the associated subset of the Grassmannian.  The two blue rows are identical, while the two red rows are identical up to an overall factor each.}\label{4Loop}
\end{center}
\end{figure}

More interesting cases arise at 3-loops and beyond where we have the possibility of tree amplitudes that are entirely internal to the loop diagram.  An example of this phenomenon is given in figure \ref{4Loop}.  Here again we have computed the kinematics of the object in twistor space by first accounting for the propagators along the boundary and then integrating over the $\WW$ variables that link the ${\cal L}^i$ along the boundary with the internal tree amplitudes.  
In the $C$ matrix structure pictured in figure \ref{4Loop}, the two blue rows are identical, while the red rows are identical up to an overall factor each.  We see again that the topology of the loop graph is reflected in the structure of the subspace of the Grassmannian.

There are many possibilities for further exploration here, and it may even be possible to categorize and understand all of the possible topologies.  Other natural goals include understanding in a more concrete way how the twistor space structure gives rise to the appropriate momentum space kinematics, and understanding whether all residues of ${\cal L}_{n,k}$ are leading singularities at all loops.  It is exciting to note that for any leading singularity our methods will give \emph{some} sub-matrix structure within ${\cal L}_{N,K}$.  Thus with one class of exceptions, we have implicitly shown that all leading singularities arise from Grassmannian kinematics.

The exceptions are the so-called ``composite leading singularities''  \cite{Cachazo:2008vp}, which seem to be important in obtaining the full loop amplitudes.  These are diagrams at $L$ loops with fewer than $4L$ explicit propagators which nevertheless give rise to leading singularities.  The classic example is the diagram
\be
\includegraphics[width=13cm]{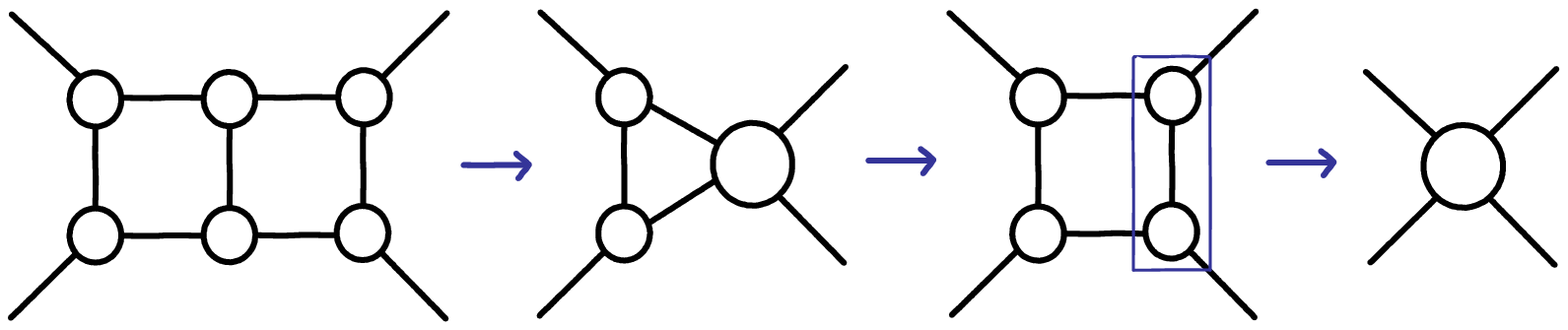} \nonumber
\ee
where we have shown the series of cuts and manipulations that one can perform in order to obtain the leading singularity.  The naive translation of this diagram into twistor space would seem not to give rise to a leading singularity, but to a product of tree amplitudes integrated over one free variable.  Clarifying the role that these sorts of leading singularities play in constructing general loop amplitudes is an important goal for future work.

\section{Conclusions and Future Diretions}

We have shown how any given leading singularity of the ${\cal N} = 4$ SYM S-Matrix can be identified among the residues of the Grassmannian contour integral $\LL_{n,k}$.  Moreover, we have seen that there is a simple and physical pattern for how the various leading singularities appear, so that the structure of the perturbation series is reflected in various subspaces within the Grassmannian.  Let us now consider some directions for future work.

\begin{itemize}

\item {\bf Evaluating the Residues}  The only piece missing from our argument is a method for calculating the residues themselves in general -- if this could be established, then our argument would become a proof that all leading singularities are residues of $\LL_{n,k}$.  Not only is this a precise mathematical problem, but we know the answer ahead of time -- for instance, at one-loop we know that there must be a residue of $\LL_{N,K}$ containing four smaller $\LL_{n_i, k_i}^i$, and we know that the actual value of the residue is given by the products of the denominator factors from the four smaller $\LL_{n_i, k_i}^i$.  However, this question remains both non-trivial and interesting, as the determinant factors that make up the denominator of $\LL_{n,k}$ make up the `Grassmannian Dynamics'.  Also, the computation of  multi-variable composite residues is in general a difficult mathematical problem \cite{GH}-\cite{Yuzhakov}, so we expect that the special form of the denominator must play a crucial role.  It will also be interesting to understand the converse statement, that all residues are in fact leading singularities, and perhaps to reverse our logic and formulate a recursive `derivation' of $\LL_{n,k}$.

\item {\bf Composite Leading Singularities}  As discussed in section 3.5, we do not have a twistor space picture for the composite leading singularites, which arise from diagrams at $L$ loops that have fewer than $4L$ explicit propagators.  These diagrams seem to play a role in the construction of the full S-Matrix \cite{Cachazo:2008vp}, so it may be important to identify them.  Another possibility is that they are somehow always associated with, or algebraically identical to, the more natural leading singularities that we have already identified.  This is a pressing issue if we hope to unite $\LL_{n,k}$ with the actual loop integrals to construct the full S-Matrix of the ${\cal N}=4$ theory\footnote{Unless of course there is some direct, once-and-for-all solution to this problem, as the existence of the Wilson Loop/Amplitude correspondence and dual conformal invariance \cite{Alday:2007hr}-\cite{McGreevy:2008zy} might be taken to suggest.}.

\item {\bf Kinematic Structures}  In our two loop $12$-pt N$^4$MHV example we saw how a particular subspace within the Grassmannian automatically encoded the solutions to the $8$ quadratic equations that arise when we `cut' $8$ loop propagators and force them on-shell -- a rather non-trivial feat.  It would be interesting to systematically understand how very complicated momentum space kinematics can be encoded by Grassmannian subspaces.  This may be of particular interest because the `Grassmannian Kinematics' may generalize beyond the ${\cal N} = 4$ theory even if the `Grassmannian Dynamics' (the denominator structure and specific residues) does not.

\item {\bf Residue Theorems}  Now that it is possible to identify leading singularities within ${\cal L}_{n,k}$ for very general $n$ and $k$ it will be interesting to try to study the appropriate residue theorems \cite{GH}-\cite{Yuzhakov} in a systematic way.  As we saw in \cite{ArkaniHamed:2009dn}, we expect that these residue theorems encode the locality of the S-Matrix by enforcing that scattering amplitudes only have physical poles and obey the IR equations.  It would be interesting to understand these facts in greater generality and at higher loops.

\item {\bf Yangian Symmetry} Although the dual conformal invariance \cite{Alday:2007hr}-\cite{McGreevy:2008zy} of ${\cal L}_{n,k}$ has been shown in \cite{ArkaniHamed:2009vw}, \cite{Mason:2009qx}, an additional miracle occurred, namely that ${\cal L}_{n,k}$ was found to be proportional to ${\cal L}_{n,k-2}$ written in a ``momentum twistor space'' \cite{Hodges:2009hk}.  In order to better understand this miracle, and also because the Yangian generators \cite{DrummondYangian} become extremely natural in twistor space, it would be interesting to directly understand the Yangian symmetry of ${\cal L}_{n,k}$.  This is not so easy because it is only the residues of ${\cal L}_{n,k}$ that are Yangian invariant; the integrand itself certainly is not.

Many of the ideas in this paper were inspired by the `Hodges diagrams' of \cite{Hodges:2005bf}-\cite{ArkaniHamed:2009si}; we have made minimal use of them mostly because they would be unfamiliar to most readers.  Previously, Hodges diagrams have only been used to represent tree amplitudes, but our method of writing leading singularities in twistor space shows that one could equally well use Hodges diagrams to represent loop-level information (in fact Hodges diagrams can enumerate all leading singularities).  It has been shown \cite{notes} that the Yangian symmetry of scattering amplitudes in the ${\cal N}=4$ theory can be seen via a simple induction argument applied to Hodges diagrams.  It would be interesting to try to extend this argument to all leading singularities.

\item {\bf Non-Supersymmetric Theories}  At one-loop, scattering amplitudes in theories such as pure Yang-Mills cannot be characterized by their leading singularities, but require the specification of so-called triangle and bubble coefficients and also rational terms that have no 4-dimensional unitarity cuts.  In 

The Hodges diagram techniques of \cite{Hodges:2005bf}-\cite{ArkaniHamed:2009si} are equally applicable to tree amplitudes in Yang-Mills theories without supersymmetry.  Using the methods of section 2, it should be possible to write triple and double cuts in twistor space, and perhaps with a bit of cleverness one could isolate the actual triangle and bubble coefficients.  Experience has shown that scattering amplitudes come back from twistor space in new and improved forms, so it might be useful to attempt to compute pure Yang-Mills amplitudes in this way.

\item {\bf Building Full Amplitudes}  It seems reasonable to interpret the very existence of $\LL_{n,k}$ as an indication of the importance of leading singularities, so it is very important to understand if there is some simple way of computing the actual S-Matrix from its leading singularities beyond one-loop.  

Another very exciting direction would involve combining the Wilson Loop, which has been conjectured to compute MHV amplitudes to all orders \cite{Drummond:2007cf}-\cite{Drummond:2008aq}, and $\LL_{n,k}$, which in the form of \cite{ArkaniHamed:2009vw} actually builds non-MHV amplitudes from MHV amplitudes using Momentum (or dual conformal) Twistors \cite{Hodges:2009hk}.  More generally, techniques from integrability \cite{Beisert:2003tq}, \cite{Bargheer:2009qu} may shed light on $\LL_{n,k}$.

\item {\bf Implications for Gravity?}  A holy grail and initial motivation for much recent work has been the hope of finding something like ${\cal L}_{n,k}$ for ${\cal N}=8$ Supergravity \cite{Cremmer:1978ds}-\cite{Cremmer:1979up}, a theory whose perturbative S-Matrix may also be determined by its leading singularities \cite{simplest}, and may be finite \cite{Bern:1998ug}-\cite{Kallosh:2008ic}.  If found, such an object could be viewed as a holographic description of flat spacetime.

The pattern of leading singularities within $\LL_{n,k}$ gives us hints for how something like ${\cal L}_{n,k}$ might work for ${\cal N}=8$ Supergravity.  Leading singularities seem to be equally important in ${\cal N}= 8$ as in ${\cal N} = 4$ \cite{simplest}, so if a direct analogue of $\LL_{n,k}$ exists for gravity, we might expect it to have the same sort of topological and recursive structure as we have found for the ${\cal N} = 4$ theory, except without color ordering.  It may make sense to ask questions along the lines of ``does there exist a manifold containing the gluing of four gravitational tree amplitudes in all possible permutations?''.  Also, we know from its non-conformal nature, from the fact that the gravitational `charge' is energy-momentum, and from explicit checks that leading singularities in ${\cal N} = 8$ cannot be characterized with as few kinematical variables as those of ${\cal N} = 4$, and this again points to a some new and different space for a dual description of gravity.

\end{itemize}

{\it Note Added:  During preparation of a companion paper to this work, an interesting new paper \cite{Bullimore:2009cb} appeared which has some overlap with this work.}

\section*{Acknowledgements}

We thank Natalia Toro and Ravi Vakil for discussions and Nima Arkani-Hamed, Freddy Cachazo, and Cliff Cheung for many relevant discussions, correspondence, and collaborations on this subject.  We also thank Jaroslav Trnka for sharing his work in progress on factorization.  We thank Jacob Bourjaily, Cliff Cheung, and Natalia Toro for comments on the draft.  We especially thank Jacob Bourjaily for many discussions, for help checking the $n=12$, $k=6$ `box-pentagon' example in momentum space, and for implementing our methods to obtain the tree level contours numerically.  JK is supported by the US DOE under contract number DE-AC02-76SF00515.

\appendix
\section{The Residues of ${\cal L}_{N,K}$}

\subsection{Jacobians}

Recall that 
\be 
{\cal L}_{n;k}({\cal W}_a) = \int \frac{d^{k \times n} C_{\alpha
a}}{(12\cdots k) \, (23\cdots (k+1) \,) \, \cdots (n 1 \cdots (k-1)
\,)} \prod_{\alpha = 1}^k \delta^{4|4}(C_{\alpha a} {\cal W}_a) 
\ee
is invariant under $GL(k)$ transformations that take $C_{\alpha a} \to L_\alpha^{\ \beta} C_{\beta a}$.  This is a redundancy of description, analogous to the gauge symmetries necessary to provide local descriptions of massless spin $1$ and spin $2$ particles (in our case the redundancy makes the cyclic permutation symmetry manifest).  This redundancy must be eliminated before we can compute leading singularities. 

We have introduced a new gauge fixing for this $GL(k)$ redundancy, so in this section we will compute the relevant Jacobian.  Perhaps the most difficult issue is coming up with a clear notation for these large matrices, so we will refer throughout to an example in the hopes that the general case is clear.  

With the `canonical' gauge fixing of \cite{ArkaniHamed:2009dn}, where the $C$ matrix is fixed so that some $k$ of its columns form the $k \times k$ identity matrix, the Jacobian is $1$.  Since our gauge fixing is very similar to this one, it will be easiest to compute our Jacobian by transforming from this gauge fixing to our own.

As a rather general example to keep in mind, a $C$ matrix with the `old' gauge fixing would be
\be
C^{\rm old} = \left(\begin{array}{cccccccccccc} 
* & * & 1 & 0 & x & 0 & x & x & 0 & 0 & * & 0 \\ 
* & * & 0 & 1 & x & 0 & x & x & 0 & 0 & x & 0 \\
* & * & 0 & 0 & * & 1 & x & x & 0 & 0 & x & 0 \\
x & x & 0 & 0 & * & 0 & * & * & 1 & 0 & x & 0 \\
x & x & 0 & 0 & x & 0 & * & * & 0 & 1 & x & 0 \\
x & x & 0 & 0 & x & 0 & * & * & 0 & 0 & * & 1
\end{array} \right) 
\ee
whereas with our gauge fixing we will take
\be
\label{OneLoopX}
C^{\rm new} = \left(\begin{array}{cccccccccccc} 
c_{11} & c_{12} & 1 & 0 & x_{15} & 0 & x_{17} & 0 & 0 & 0 & c_{1,11} & c_{1,12} \\ 
c_{21} & c_{22} & c_{23} & 1 & x_{25} & 0 & x_{27} & x_{28} & 0 & 0 & 0 & 0 \\
c_{31} & c_{32} & c_{33} & 0 & c_{35} & 1 & x_{37} & x_{38} & 0 & 0 & 0 & 0 \\
x_{41} & 0 & 0 & 0 & c_{45} & c_{46} & c_{47} & c_{48} & 1 & 0 & x_{4,11} & 0 \\
x_{51} & x_{52} & 0 & 0 & 0 & 0 & c_{57} & c_{58} & c_{59} & 1 & x_{5,11} & 0 \\
x_{61} & x_{62} & 0 & 0 & 0 & 0 & c_{67} & c_{68} & c_{69} & 0 & c_{6,11} & 1
\end{array} \right) 
\ee
It is easy to find the $GL(k)$ transformation that relates these two matrices.  We simply take
\be
C^{\rm new}_{\alpha a} = J \left( C^{\rm new} \right)_\alpha^{\ \beta} C^{\rm old}_{\beta a} \ \ \ {\rm with} \ \ \ J \left( C^{\rm new} \right) = \left(\begin{array}{cccccc} 
1 & 0 & 0 & 0 & 0 & c_{1,11}  \\
c_{23} & 1 & 0 & 0 & 0 & 0  \\
c_{33} & 0 & 1 & 0 & 0 & 0  \\
0 & 0 & c_{46} & 1 & 0 & 0  \\
0 & 0 & 0 & c_{59} & 1 & 0  \\
0 & 0 & 0 & c_{69} & 0 & 1  
\end{array} \right) 
\ee
We have emphasized that $J$ is a function of the $C^{\rm new}$ variables, so that $J^{-1} \cdot C^{\rm new}$ also depends entirely on these variables.
Now we can compute the Jacobian from the equation $J^{-1} \cdot C^{\rm new} = C^{\rm old}$.  Taking $d$ of both sides and then multiplying by $J$ gives
\be
d C^{\rm new}_{\alpha a} + (J \cdot dJ^{-1})_{\alpha}^{\ \beta} \cdot C^{\rm new}_{\beta a} = J_{\alpha}^{\ \beta} \cdot d C^{\rm old}_{\beta a}
\ee
Since ${\cal L}_{n,k}$ is invariant under global $GL(k)$ transformations, this last multiplication with $J$ drops out of the overall Jacobian, which we can now compute directly from the left hand side of the equation above.  It is amusing that this equation makes it manifest that $J$ is a $GL(k)$ ``gauge field''.  Also note that this equation is completely general, and does not depend on any of the details of our particular illustrative example.

Now the measure comes from taking the wedge product
\be
\bigwedge_{\alpha, a} \left[d C^{\rm new}_{\alpha a} + (J \cdot dJ^{-1})_{\alpha}^{\ \beta} \cdot C^{\rm new}_{\beta a} \right]
\ee
and the variables that do not appear in $J$ can be factored out.  This means that the only columns (values of $a$) that produce a non-trivial Jacobian are those where there are extra $0$s in $C^{\rm new}$.  Thus the Jacobian is
\be
\prod_{i=1}^K \left( J \cdot \frac{\partial J^{-1}}{\partial c_i} \right)_{\alpha_i}^{\beta} C_{\beta a_i}^{\rm new}
\ee
where $i$ labels the $K$ entries in $C^{\rm new}$ that have been set to zero by our gauge fixing, and $a_i$ and $\alpha_i$ are the corresponding columns and rows.  This formula simply reduces to a product of minors to various powers; in the case of our example the Jacobian is
\be
(c_{32} - c_{12} c_{33}) (c_{45} - c_{35} c_{46})^2 (c_{68} - c_{48} c_{69}) (c_{1,11} - c_{6,11} c_{1,12})^2
\ee
In general, with our specific gauge fixing, the Jacobian is given by a product of four $(k_i - 1) \times (k_i - 1)$ minor determinants, each raised to the power $k_{i+1} - 1$.  These are the right-most minors in each of the $(n_i - 2) \times (k_i - 1)$ sub-blocks corresponding to the four corners of the one-loop leading singularity (box), as can be verified by a straightforward computation.

\subsection{Existence of Tree and One-Loop Residues}

In this appendix we will argue for the existence of the residues of ${\cal L}_{N,K}$ that give rise to the block structure of the $C$ matrix corresponding to the four ${\cal L}^i$.  We will refer to the Grassmannian coordinates that we wish to set to zero as $x$ variables, as pictured in the $C$ matrix of equation (\ref{OneLoopX}).  We will show that the denominator of ${\cal L}_{N,K}$ vanishes to high enough order in the $x$ variables for the point $x=0$ to be a residue.

To begin let us count the number of $x$ variables, noting for convenience that $N + 8 = \sum_i n_i$ and $K + 4 = \sum_i k_i $.  There are $NK - \sum k_i (n_i - 2)$ entries in $C$ outside of the sub-matrices corresponding to the ${\cal L}^i$, but $K + \sum (k_i-1) (K-k_i)$ are set to zero once we fix the $GL(K)$ redundancy, so there are 
\be
N_x = (N-K)K + 4 - \sum_{i=1}^4 \left[ k_i(n_i - k_i) \right]
\ee
$x$ variables in total.  Now we need to show that 
\be
D = (12...K)(23...K+1)...(N12...K-1)
\ee
has no terms of lower order lower than this in the $x$ variables.  Another way of saying this is that we want to prove that the denominator, considered as a polynomial in the $x$ variables, is to leading order homogeneous and of degree $N_x$.

It suffices to examine how the rank of the $K \times K$ matrices appearing in $D$ depends on the $x$ variables.  Specifically, we would like to consider how the sum of the ranks of these $N$ matrices changes when $x$ take generic values versus when all $x=0$, since this tells us the order of $D$ as a polynomial in the $x$.  For each $i=1,2,3,4$ there are $k_i-2$ rows full of $x$s (or zeroes) that are each of length $N + 2 - n_i$ and also four rows of length $N + 4 - n_i - n_{i+1}$.  The presence of each row increases the order of $D$ in the $x$ variables by the length of the row minus $K-1$.  However, there is an additional effect near the corners of the ${\cal L}^i$ sub-matrices because a linear dependence in either the rows or the columns of a matrix will decrease its rank.  This contributes $(k_i-1)(k_i-2)/2$ at two corners of each of the four sub-matrices, giving a total
\be
(N-K+3)K + 8 + \sum_{i=1}^4 \left[ (k_i-1)(k_i-2) - k_i n_i \right]
\ee
This is precisely equal to the number of $x$ variables $N_x$ that we counted above.  Without a better understanding of the precise definition of the residue we cannot conclude that it exists, but our argument makes it very plausible.

\section{All NMHV Residues}

Now we will give a solution for all the residues of ${\cal L}_{n,3}$.  By a solution we mean an explicit identification of every residue of the contour integral
\be 
{\cal L}_{n;3}({\cal W}_a) = \int \frac{d^{3n} C_{\alpha
a}}{(123) \cdots (i-1, i, i+1) \cdots (n 1 2)} \prod_{\alpha = 1}^3 \delta^{4|4}(C_{\alpha a} {\cal W}_a) 
\ee
This is a multi-dimensional contour integral over a $G(3,n)$ Grassmannian; it is useful to count the number of integration variables in order to see the best way to label the residues.  After eliminating the $GL(3)$ redundancy of the Grassmannian, ${\cal L}_{n,3}$ becomes an integral over $3n-9$ variables.  When we Fourier transform from twistor space back to momentum space, we produce $2n$ delta functions, but $4$ of these turn into the momentum conservation delta function, so there are only $2n-4$ independent constraints.  After these constraints have been taken into account ${\cal L}_{n,3}$ reduces to a contour integral over $(3n-9) - (2n-4) = n-5$ free variables.  The denominator of the integrand is simply a product of $n$ $3 \times 3$ determinants, and on the delta function constraints these are each linear functions of the $n-5$ free variables.  Thus a single residue can be specified by listing the $5$ determinants that are \emph{not} set to zero at the residue of the contour of integration.  

It is easiest to think of the solution as being given by this diagram
\be
\includegraphics[width=9cm]{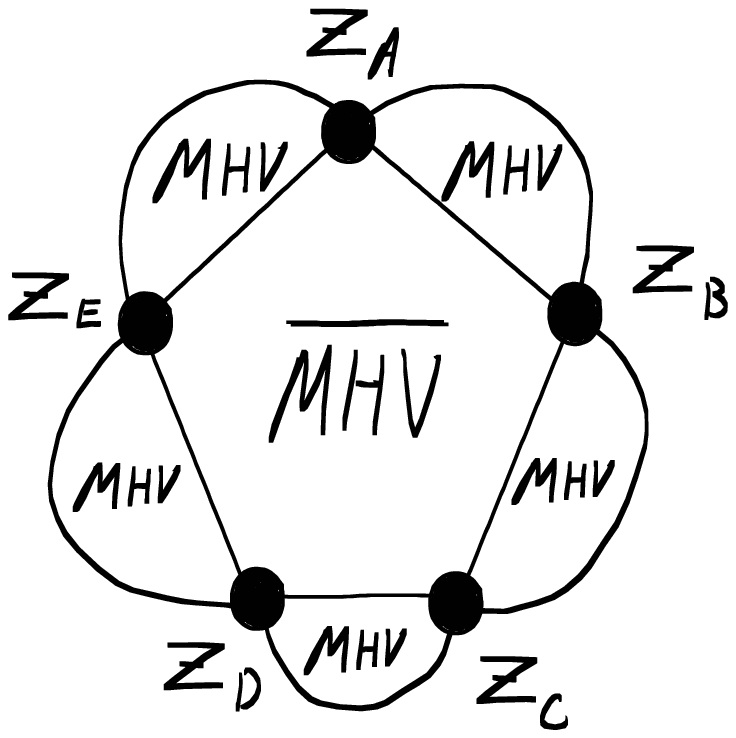} \nonumber
\ee
Those familiar with \cite{ArkaniHamed:2009si} may note that this is a `Hodges diagram', but knowledge of these diagrams is not essential to understand what follows.  The particles are labelled by an integer from $1$ to $n$, and $A,B,C,D,E$ can be any increasing set of integers in this range.  We are representing these particles at the vertices of the pentagon with $\ZZ = (\tilde \lambda, \tilde \mu, \tilde \eta)$ twistors, while all of the other particles, which are not explicitly drawn, are most naturally taken to be $\WW = (\lambda, \mu, \eta)$ twistors.  This is simply a choice of basis and is not physically meaningful, but it will be useful in what follows.  What the diagram means is that we take the anti-MHV 5-pt amplitude $\MM_5(\ZZ_A,\ZZ_B,\ZZ_C,\ZZ_D, \ZZ_E)$ and up to five MHV amplitudes such as $\MM(\ZZ_A, \WW_{A+1},..., \ZZ_B)$ and \emph{simply multiply them}.  The number of particles in each MHV amplitude is fixed by differences such as $B-A$; if $B = A+1$ then there is no MHV amplitude on the $AB$ side of the pentagon.

What does the diagram mean physically?  It turns out that this is the most general object that one can get from applying the BCFW recursion relations to compute NMHV tree amplitudes.  Our claim is that this diagram is precisely the residue that we would label $\{A, B, C, D, E \}$, where eg $A$ represents the determinant $(A-1, A, A+1)$.  Let us now show this explicitly.

First we will write ${\cal L}_{n,3}$ in a basis where particles $A, B, C, D, E$ are represented by $\ZZ$ and the others are represented by $\WW$ in order to facilitate comparison with the diagram.  We fix the $GL(3)$ redundancy of ${\cal L}_{n,3}$ by setting columns $A$, $B$, and $C$ to the identity matrix.  Next we Fourier transform these particles to the $\ZZ$ basis, giving
\be 
{\cal L}_{n;3} = \int \frac{d^{3n-9} c_{i
J}}{(123) \cdots (i-1, i, i+1) \cdots (n 1 2)} e^{i c_{i J} \WW_i \cdot \ZZ_J}
\ee
where $J = A,B,C$.  Now we can Fourier transform particles $D$ and $E$ to the $\ZZ$ basis as well, giving
\be 
{\cal L}_{n;3} = \int \frac{d^{3n-9} c_{i
J}}{(123) \cdots (i-1, i, i+1) \cdots (n 1 2)} e^{i c_{i J} \WW_i \cdot \ZZ_J}
\delta^{4|4}(Z_D + c_{D J} Z_J) \delta^{4|4}(Z_E + c_{E J} Z_J)
\ee
The residue of interest is obtained by setting $(I-1,I,I+1)=0$ for all $I \neq A,B,C,D,E$.  We will now see that the diagram can be written as an integral over the same $c_{iJ}$ variables with the same structure of delta functions as ${\cal L}_{n,3}$.  The vanishing of the claimed determinants will be guaranteed by the structure of the diagram.

The central pentagon of the diagram is simply an anti-MHV $5$-pt amplitude.  In accord with our choice of variables for ${\cal L}_{n,3}$ let us represent it in the all $\ZZ$ basis with its $C_{\alpha a}$ matrix fixed to be
\be
C^p = \left(\begin{array}{ccccc} c^p_{DA} & c^p_{DB} & c^p_{DC} & 1 & 0 \\ c^p_{EA} & c^p_{EB} & c^p_{EC} & 0 & 1 \end{array} \right)
\ee
where the $p$ index indicates that these are the $c$'s in the pentagon.  Now we can write the pentagon as
\be
\int \frac{d^6 c^p}{(AB)(BC)...(EA)} \delta^{4|4}(\ZZ_D + c^p_{DJ} \ZZ_J) \delta^{4|4}(\ZZ_E + c^p_{EJ} \ZZ_J)
\ee
and the delta functions match up with our representation of ${\cal L}_{n;3}$.  

Our diagram represents the product of this pentagon with the five MHV amplitudes that are attached to its edges.  We can write each of these amplitudes as a copy of ${\cal L}_{m;2}$.  For example, the MHV amplitude attached to $A$ and $B$ can be written with a $C^{AB}$ matrix
\be
C^{AB} = \left(\begin{array}{cccccc} 1 & c_{A+1, A} & c_{A+2,A} & ...  & c_{B-1, A} & 0 \\ 0 & c_{A+1, B} & c_{A+2, B} & ... & c_{B-1, B} & 1 \end{array} \right)
\ee
so that the MHV amplitude itself takes the form
\be
\MM^{DE} = \int \frac{d C^{AB}}{(A, A+1)... (B-1, B)(B, A)} e^{i c_{iJ}^{AB} \WW_i \cdot \ZZ_J}
\ee
where $J = A, B$ and $i$ runs from $A+1$ to $B-1$.  Similar expressions obtain for  $\MM^{BC}, \MM^{CD}, \MM^{DE}$, and $\MM^{EA}$.  

Now we can see that with our choice of ``gauge fixing'' of the various $GL(2)$ and $GL(3)$ Grassmannian redundancies of description, $(I-1, I, I+1) = 0$ for $A < I < B$ but that $(B-1, B, B+1)$ does not vanish.  The former statement follows from the fact that $c_{I,C}$ does not exist, so in other words  $c_{I,C} = 0$ by definition.  Since the determinant factors are linear in $c_{I,C}$, they vanish.  The latter statement follows by direct evaluation -- $(B-1, B, B+1) = c_{B-1, A} c_{B+1, C}$ which can be seen to be non-vanishing in momentum space by a direct computation.

Both the diagram and ${\cal L}_{n,3}$ are independent of the choice of $\ZZ$ or $\WW$ basis and the ``gauge fixing'' of the various $GL(2)$ and $GL(3)$ redundancies.  With different gauge fixings it would be clear that the determinant $(I-1,I,I+1) = 0$ for all $I \neq A,B,C,D,E$.  Since we have made no assumptions that break the symmetry between $A, B, C, D, E$ except for the choice of basis and ``gauge'', we can conclude that the diagram corresponds to the claimed residue of ${\cal L}_{n,3}$.

\end{document}